\documentclass[journal]{IEEEtran}
\usepackage{amsmath}
\usepackage{amssymb}
\usepackage{algorithm}
\usepackage{graphicx}
\usepackage{algpseudocode}
\usepackage{booktabs}

\begin{document}

\title{Time Segmented Beamforming via Dynamic Programming: Theory and Implementation}

\author{Manan~Mittal,
        Ryan~M.~Corey,
        Diego~Cuji,
        John~R.~Buck,
        and~Andrew~C.~Singer%
\thanks{M. Mittal and D. Cuji are with the Department of Electrical and Computer Engineering, Stony Brook University, Stony Brook, NY 11794 USA (e-mail: manan.mittal@stonybrook.edu).}%
\thanks{R. M. Corey is with the Department of Electrical and Computer Engineering, University of Illinois, Chicago, IL 60607 USA}%
\thanks{J. R. Buck is with the Department of Electrical and Computer Engineering, University of Massachusetts Dartmouth, Dartmouth, MA 02747 USA.}%
\thanks{A. C. Singer is with the College of Applied Science and Engineering, Stony Brook University, Stony Brook, NY 11794 USA.}%
}

\maketitle

\begin{abstract}
In dynamic acoustic environments with time-varying interferers, effective beamforming requires identifying stationary regions over time. The Capon beamformer, a whitened matched filter constrained to maintain unity gain in the desired direction, theoretically relies on the instantaneous ensemble covariance matrix. Practical implementations rely on the batch Capon (or Sample Matrix Inversion), which estimates the sample covariance matrix (SCM) by averaging over a block of snapshots. This practical approach implicitly assumes that the data within the batch window is stationary and can be coherently combined. In non-stationary settings, a batch approach that averages over fixed or excessively long windows fails, as moving interferers smear the SCM and degrade the beamformer's nulling capabilities. To address this, this paper introduces a temporally segmented distortionless response beamformer. Inspired by the segmented least squares method, which fits piecewise polynomials to data while penalizing excessive segmentation to prevent overfitting, the framework extends practical Capon beamforming by incorporating data-driven temporal segmentation. This formulation minimizes output power while dynamically adapting the SCM estimation windows to local stationarity, offering a principled approach to tracking time-varying interferers.
\end{abstract}



\section{Introduction}

Beamforming algorithms function as spatial filters, processing measurements from a distributed sensor array to enhance signals arriving from a specific direction while suppressing interference and noise. By applying direction-dependent weights to each sensor channel, the beamformer coherently combines the desired signal components, often modeled via a steering vector or relative transfer function (RTF), and destructively interferes with those originating from other spatial locations. This process improves the signal-to-noise ratio (SNR) of the target source, making it a fundamental technique in acoustic signal processing, radar, and wireless communications.

The standard benchmark for adaptive spatial filtering is Capon’s Minimum Variance Distortionless Response (MVDR) beamformer \cite{capon}. The MVDR formulation seeks a weight vector that minimizes the total output power of the array, subject to a linear constraint that maintains a unity response in the look direction of the desired signal. By strictly constraining the target response, the algorithm ensures that any minimization of the total output power is achieved exclusively through the suppression of noise and interfering signals. Theoretically, this solution relies on the ensemble covariance matrix (ECM) \cite{van_trees_optimum_2002}, which represents the true second-order statistics of the received data. Therefore, the Capon beamformer provides the a distortionless estimate of the source signal and achieves the lower bound on output power for a linear filter.

The practical utility of the Capon beamformer faces constraints because the system must estimate the covariance matrix from finite data. This estimation introduces a fundamental bias-variance tradeoff governed by the integration window length. To minimize estimation variance and ensure the sample covariance matrix (SCM) is well-conditioned, a long integration window is required. Conversely, physical acoustic environments exhibit non-stationary behaviors; interferers appear and vanish, sources move, and reverberation patterns evolve. In such dynamic settings, a long integration window spans multiple distinct acoustic states. The resulting SCM represents a smeared temporal average that does not correspond to any singular physical state of the environment. Consequently, a beamformer derived from this average places nulls in directions where interferers used to be, while simultaneously failing to track newly active interference sources. This mismatch wastes the spatial degrees of freedom of the array and degrades signal recovery.

Traditional approaches to non-stationary beamforming typically address this issue through fixed-memory mechanisms, such as sliding windows, exponential forgetting factors, or multi-rate adaptive processing techniques \cite{cox_multirate}. While these methods allow for adaptation, they rely on hyperparameters that implicitly assume a specific, constant timescale of stationarity. This assumption is frequently violated in practice, as the rate of environmental change is unknown and time-varying. A fixed window optimized for a slowly moving source may fail to track a sudden onset of interference, leading to performance degradation during state transitions.

To address the limitations of fixed-memory estimation, this work reformulates the beamforming problem as a joint task of estimation and temporal segmentation. The framework posits that the optimal integration time should function as a dynamic variable derived directly from the statistical structure of the data. By applying the principles of dynamic programming, specifically Bellman’s Segmented Least Squares (SLS) framework \cite{bellman, bellman1961approximation}, this paper introduces the Batch Segmented Beamformer (BSB). This method partitions the observation record into variable-length segments, within which the data is approximated as stationary.

The core contribution of this framework is the derivation of a global objective function that minimizes the cumulative output power across all segments, penalized by a cost term for model complexity. This penalty prevents overfitting, a condition where every sample forms an independent segment, and enforces a parsimonious representation of the acoustic scene. Unlike standard adaptive methods that uniformly update weights at every step, the BSB identifies the optimal boundaries where the acoustic statistics shift, effectively generalizing the classical Capon estimator to piecewise-stationary environments.

While the batch formulation of the Segmented Beamformer provides a globally optimal solution, it is inherently non-causal. To enable real-time processing, the paper details an Online Segmented Beamformer (OSB) \cite{mittal2026online}, inspired by the Online Segmented Recursive Least Squares (OSRLS) algorithm \cite{osrls}. This causal approximation operates sequentially, evaluating at each time step whether to extend the current stationary segment or to declare a change point and reset the covariance estimate. This approach allows the beamformer to adapt instantaneously to abrupt changes in the interference environment without the lag traditionally associated with sliding windows. The theoretical contribution encompasses a regret bound demonstrating that the online algorithm achieves vanishing regret at a logarithmic rate compared with the batch segmented beamformer.

The evaluation validates the proposed framework through simulations and experimental analysis on the SwellEx-96 dataset and a distributed microphone array. The results demonstrate that by explicitly modeling the temporal structure of the interference, the OSB achieves lower output power and higher interference suppression than fixed-window techniques, effectively adapting its effective memory length to the dynamics of the environment.

\section{Signal Model}
Consider an array comprising $N_m$ sensors and $N_s$ acoustic sources. The sensors are arbitrarily distributed with no specific geometric constraints, and the architecture assumes a centralized processing pipeline where all sensor measurements are jointly available.

Let $s_j$ denote the signal from the $j^{\text{th}}$ source. The signal received at sensor $m$ is the convolution of $s_j$ with the acoustic transfer function $h_{m,j}$, plus additive spatially uncorrelated noise $v_m$. The observed signal $x_m[t]$ at time $t$ takes the form:
\begin{equation}
x_m[t] = \sum_{j=1}^{N_s} (h_{m,j} \ast s_j)[t] + v_m[t],
\end{equation}
where $\ast$ denotes convolution.

The beamformer applies a linear filter $w_m$ to each sensor channel and sums the resulting outputs. The beamformed output $z[t]$ takes the form:
\begin{equation}
z[t] = \sum_{m=1}^{N_m} (w_m \ast x_m)[t].
\end{equation}

To facilitate analysis, this operation uses vector notation. The formulation assumes each filter $w_m$ is a finite impulse response (FIR) filter of length $L$. Let $p = N_m L$ denote the total spatio-temporal dimension of the processing space. The coefficients populate a single time-varying weight vector $\mathbf{w}[t] \in \mathbb{R}^{p}$:
\begin{equation*}
\mathbf{w}[t] = \begin{bmatrix} w_1[t,0], \dots, w_1[t,L{-}1], \dots,\\ w_{N_m}[t,0], \dots, w_{N_m}[t,L{-}1] \end{bmatrix}^\top.
\end{equation*}
Similarly, the system concatenates the most recent $L$ samples from all sensors into the observation vector $\mathbf{x}[t] \in \mathbb{R}^{p}$:
\begin{equation*}
\mathbf{x}[t] = \begin{bmatrix} x_1[t], \dots, x_1[t{-}L{+}1], \dots, x_{N_m}[t], \dots, x_{N_m}[t{-}L{+}1] \end{bmatrix}^\top.
\end{equation*}
The beamformer output reduces to the inner product:
\begin{equation}
z[t] = \mathbf{w}[t]^\top \mathbf{x}[t].
\end{equation}

The model assumes all signals are real-valued and zero-mean. The instantaneous second-order statistics define the correlation matrix $\mathbf{R}[t] = \mathbb{E}[\mathbf{x}[t] \mathbf{x}[t]^\top] \in \mathbb{R}^{p \times p}$, where $\mathbb{E}[\cdot]$ denotes expectation.

In many applications, partial knowledge of the desired source's spatial signature is available. For example, the conventional beamformer weights, which optimize white-noise gain, correspond to the steering vector in plane-wave environments. In reverberant or unstructured environments where analytic steering vectors are unavailable, the system estimates the Relative Transfer Function (RTF) to characterize the relative gains and delays of the desired source.

The OSB applies to both cases. Whether the steering vector is known a priori or estimated as an RTF, the resulting constraint vector corresponds to $\boldsymbol{\nu} \in \mathbb{R}^p$ \cite{consolidated}.

\section{Capon Beamformer}
\subsection{Ensemble Beamformer}
The Capon beamformer minimizes the expected output power of the array while maintaining a distortionless response in the look direction. This formulation operates as a constrained matched filter that adaptively places nulls to suppress interference and noise.

Mathematically, the optimal weight vector $\mathbf{w}_{\text{opt}}$ solves the constrained optimization problem:
\begin{equation}
    \min_{\mathbf{w}} \quad J(\mathbf{w}) = \mathbb{E}\big[ \mathbf{w}^\top \mathbf{R}[t] \mathbf{w} \big] \quad \text{subject to} \quad \mathbf{w}^\top \boldsymbol{\nu} = 1,
\end{equation}
where $\mathbb{E}[\cdot]$ denotes the expectation operator and $\mathbf{R}[t] = \mathbb{E}[\mathbf{x}[t]\mathbf{x}[t]^\top]$ represents the ensemble covariance matrix.

This convex optimization problem solves analytically using the method of Lagrange multipliers. The Lagrangian function takes the form:
\begin{equation}
    \mathcal{L}(\mathbf{w}, \lambda) = \mathbf{w}^\top \mathbf{R}[t] \mathbf{w} + \lambda (1 - \mathbf{w}^\top \boldsymbol{\nu}),
\end{equation}
where $\lambda$ represents the Lagrange multiplier. Taking the gradient with respect to $\mathbf{w}$, setting it to zero, and enforcing the linear constraint yields the classic MVDR solution:
\begin{equation}
    \mathbf{w}_{\text{opt}}[t] = \frac{\mathbf{R}[t]^{-1}\boldsymbol{\nu}}{\boldsymbol{\nu}^\top \mathbf{R}[t]^{-1}\boldsymbol{\nu}}.
\end{equation}

This solution establishes the optimal spatial filter for a given covariance structure. If the ensemble statistics of the acoustic environment were known a priori, a system designer could implement these weights directly. However, in practical real-time environments, the true covariance matrix $\mathbf{R}[t]$ remains unknown and requires estimation from the finite set of received snapshots $\mathbf{x}[t]$.

\subsection{Adaptive MVDR Beamformer}
In practical implementations, the system estimates the beamforming weights adaptively from the received data sequence. The formulation assumes that the steering vector $\boldsymbol{\nu}$ corresponding to the desired signal is known or estimated a priori. The sample covariance matrix (SCM), denoted $\mathbf{S}[t]$, approximates the ensemble covariance computed over a window of $K$ snapshots:
\begin{equation}
    \mathbf{S}[t] = \frac{1}{K} \sum_{n=t-K}^{t-1} \mathbf{x}[n]\mathbf{x}[n]^\top.
\end{equation}
A standard approach employs the growing window model ($K=t-1$), which utilizes the entire history of available data. For real-time applications, the weight update requires strictly causal operations; the SCM at time $t$ relies solely on the snapshot history $\mathbf{x}[1], \dots, \mathbf{x}[t-1]$ and cannot utilize the current observation $\mathbf{x}[t]$. Consequently, the adaptive MVDR weights at time $t$ become:
\begin{equation}
    \mathbf{w}[t] = \frac{\mathbf{S}[t]^{-1}\boldsymbol{\nu}}{\boldsymbol{\nu}^\top \mathbf{S}[t]^{-1}\boldsymbol{\nu}}.
\end{equation}

Algorithm \ref{alg:adaptive_MVDR} details the efficient implementation of the adaptive MVDR beamformer. To minimize computational cost, the procedure employs the Woodbury matrix identity to update the inverse covariance matrix recursively. This allows the optimal weights to process via a rank-one update from the previous time step, avoiding direct matrix inversion. While this formulation uses a growing window, exponential forgetting factors can integrate easily to weight recent data more heavily, analogous to Recursive Least Squares (RLS) estimation.

\begin{algorithm}[ht!]
\caption{Adaptive MVDR beamformer}\label{alg:adaptive_MVDR}
\begin{algorithmic}[1]
\Require $\mathbf{x}[t] \in \mathbb{R}^p$ : Received signal snapshot at time $t$
\Require $\boldsymbol{\nu} \in \mathbb{R}^p$ : Steering vector
\Require $\delta$ : Diagonal loading factor
\Ensure $\mathbf{S}^{-1}[t]$ : Inverse sample covariance matrix
\Ensure $\mathbf{P}[t]$ : Weight numerator state vector
\Ensure $\mathbf{w}[t]$ : MVDR beamformer weight vector
\Ensure $z[t]$ : Filtered output at time $t$

\State \textbf{Initialization:}
\State $\mathbf{S}^{-1}[0] = \delta^{-1} \mathbf{I}_{p}$
\State $\mathbf{P}[0] = \delta^{-1} \boldsymbol{\nu}$
\State $\mathbf{w}[0] = \boldsymbol{\nu} / (\boldsymbol{\nu}^\top \boldsymbol{\nu})$
\State $t = 1$

\While{$t \leq T$}
    \State \textbf{Compute filtered output:}
    \State \quad $z[t] = \mathbf{w}[t-1]^\top \mathbf{x}[t]$
    
    \State \textbf{Update Filter State:}
    \State \quad $\mathbf{u}[t] = \mathbf{S}^{-1}[t-1] \mathbf{x}[t]$ \Comment{Compute intermediate vector}
    \State \quad $\gamma[t] = 1 + \mathbf{x}[t]^\top \mathbf{u}[t]$ \Comment{Compute scalar normalization factor}
    \State \quad $\mathbf{S}^{-1}[t] = \mathbf{S}^{-1}[t-1] - \frac{\mathbf{u}[t] \mathbf{u}[t]^\top}{\gamma[t]}$ \Comment{Woodbury update of inverse covariance}
    \State \quad $\mathbf{P}[t] = \mathbf{P}[t-1] - \frac{\mathbf{u}[t] (\mathbf{u}[t]^\top \boldsymbol{\nu})}{\gamma[t]}$ \Comment{Update weight numerator}
    \State \quad $\mathbf{w}[t] = \frac{\mathbf{P}[t]}{\boldsymbol{\nu}^\top \mathbf{P}[t]}$ \Comment{Update beamforming weights}
    
    \State \textbf{Increment time index:}
    \State \quad $t = t + 1$
\EndWhile
\end{algorithmic}
\end{algorithm}

\subsection{Generalized Sidelobe Canceler}
Many practical implementations of the MVDR beamformer utilize the Generalized Sidelobe Canceler (GSC) structure. The GSC operates on the principle of subspace decomposition, splitting the weight vector into two orthogonal components: a non-adaptive component that satisfies the linear constraints, and an adaptive component that minimizes output power within the unconstrained subspace.

The quiescent branch projects the received signal onto the constraint subspace defined by the steering vector $\boldsymbol{\nu}$, effectively computing the output of a conventional delay-and-sum beamformer. The adaptive branch processes the signal through a blocking matrix $\mathbf{B}$, which projects the data onto the null space of the constraints ($\mathbf{B}^\top \boldsymbol{\nu} = 0$). This ensures that the desired signal remains blocked from the adaptive path, preventing target cancellation.

Formally, the formulation decomposes the beamforming weight vector as:
\begin{equation}
    \mathbf{w} = \mathbf{w}_q - \mathbf{B}\mathbf{w}_a,
\end{equation}
where $\mathbf{w}_q$ defines the quiescent weight vector, $\mathbf{B}$ serves as the blocking matrix, and $\mathbf{w}_a$ represents the unconstrained adaptive filter. The beamformer output $z[t]$ calculates as:
\begin{equation}
    \begin{split}
        z[t] = \mathbf{w}^\top \mathbf{x}[t] &= \mathbf{w}_q^\top \mathbf{x}[t] - \mathbf{w}_a^\top \mathbf{B}^\top \mathbf{x}[t] \\
        &= d[t] - \mathbf{w}_a^\top \mathbf{u}[t],
    \end{split}
\end{equation}
where $d[t]$ denotes the desired response (quiescent output) and $\mathbf{u}[t] = \mathbf{B}^\top \mathbf{x}[t]$ denotes the interference reference signal (blocking matrix output).

In this formulation, the constrained minimization of the output power transforms into an unconstrained linear least-squares estimation problem. The adaptive filter $\mathbf{w}_a$ attempts to estimate the noise component present in $d[t]$ using the reference $\mathbf{u}[t]$. The optimization objective takes the form:
\begin{equation}
\begin{split}
    J(\mathbf{w}_a) &= \mathbb{E}\big[ |d[t] - \mathbf{w}_a^\top \mathbf{u}[t]|^2 \big].
\end{split}    
\end{equation}
Minimizing this mean-squared error equates mathematically to the standard MVDR problem. 

Algorithm \ref{alg:causal_GSC} provides an implementation of the adaptive GSC using a growing covariance window. The adaptive weights update recursively using the Woodbury matrix identity, similar to the Recursive Least Squares (RLS) algorithm.

\begin{algorithm}
\caption{Causal GSC with Unity Gain Constraint}\label{alg:causal_GSC}
\begin{algorithmic}[1]
\Require $\mathbf{x}[t] \in \mathbb{R}^p$ : Received signal snapshot at time $t$
\Require $\boldsymbol{\nu} \in \mathbb{R}^p$ : Steering vector
\Require $\delta$ : Diagonal loading factor
\Ensure $\mathbf{S}^{-1}[t]$ : Inverse sample covariance of adaptive branch
\Ensure $\mathbf{w}_a[t]$ : Adaptive sidelobe cancellation weights
\Ensure $z[t]$ : Filtered output at time $t$
\Ensure $d[t]$ : Quiescent output at time $t$

\State \textbf{Initialization:}
\State $\mathbf{S}^{-1}[0] = \delta^{-1} \mathbf{I}_{p}$
\State $\mathbf{w}_q = \boldsymbol{\nu} (\boldsymbol{\nu}^\top \boldsymbol{\nu})^{-1}$ \Comment{Quiescent weights (Moore-Penrose)}
\State $\mathbf{w}_a[0] = \mathbf{0}$ \Comment{Initial adaptive weights}
\State $\mathbf{B} = \mathbf{I}_{p} - \boldsymbol{\nu}(\boldsymbol{\nu}^\top\boldsymbol{\nu})^{-1}\boldsymbol{\nu}^\top$ \Comment{Blocking Matrix (Null space projector)}
\State $t = 1$

\While{$t \leq T$}
    \State \textbf{Compute Branch Outputs:}
    \State \quad $d[t] = \mathbf{w}_q^\top \mathbf{x}[t]$ \Comment{Quiescent path}
    \State \quad $\mathbf{u}[t] = \mathbf{B}^\top \mathbf{x}[t]$ \Comment{Adaptive path input (Blocked signal)}
    \State \quad $z[t] = d[t] - \mathbf{w}_a[t-1]^\top \mathbf{u}[t]$ \Comment{Final beamformer output}

    \State \textbf{Update Inverse Covariance (RLS):}
    \State \quad $\mathbf{k}[t] = \mathbf{S}^{-1}[t-1] \mathbf{u}[t]$ \Comment{Compute gain precursor}
    \State \quad $\gamma[t] = 1 + \mathbf{u}[t]^\top \mathbf{k}[t]$ \Comment{Compute scalar normalization}
    \State \quad $\mathbf{S}^{-1}[t] = \mathbf{S}^{-1}[t-1] - \frac{\mathbf{k}[t] \mathbf{k}[t]^\top}{\gamma[t]}$ \Comment{Woodbury update}

    \State \textbf{Update Adaptive Weights:}
    \State \quad $\mathbf{w}_a[t] = \mathbf{w}_a[t-1] + \frac{\mathbf{k}[t] z[t]}{\gamma[t]}$ \Comment{Weight update}

    \State \textbf{Increment Time Index:}
    \State \quad $t = t + 1$
\EndWhile
\end{algorithmic}
\end{algorithm}

\section{Segmented Least Squares}
\subsection{Piecewise Approximation of Curves using Line Segments}

In foundational work on the approximation of curves \cite{bellman1961approximation}, Bellman addressed the following problem: Given a signal $x(t)$ defined on an interval $[a,b]$, determine the optimal partitioning of $x(t)$ into $N$ linear segments to minimize the cumulative squared error.

Let the $N+1$ points defining the boundaries of these partitions correspond to $a = \tau_0 < \tau_1 < \dots < \tau_N = b$. First, the error of approximating the signal with a single line segment over the full interval $[a, b]$ defines the baseline error, denoted $E_1(b)$, given by:
\begin{equation}
    E_1(b) = \min_{m, c} \int_a^b (x(t) - m t - c)^2 dt,
\end{equation}
where $m$ and $c$ represent the slope and intercept of the linear estimator.

Extending this to two segments involves finding an intermediate boundary $\tau_1$ that minimizes the sum of the errors of the two distinct linear fits. The two-segment error $E_2(b)$ takes the form:
\begin{equation}
    \begin{split}
        E_2(b) = \min_{\tau_1} \Bigg[ &\min_{m_1, c_1} \int_a^{\tau_1} (x(t) - m_1 t - c_1)^2 dt \\
        &+ \min_{m_2, c_2} \int_{\tau_1}^b (x(t) - m_2 t - c_2)^2 dt \Bigg].
    \end{split}
\end{equation}
The optimal boundary $\tau_1$ resolves via a one-dimensional search over the interval $[a, b]$.

When generalizing to $N$ segments, the problem exhibits optimal substructure. The cumulative error $E_N(b)$ expresses recursively using Bellman’s Principle of Optimality \cite{bellman}. The error for $N$ segments represents the minimum sum of the error for $N-1$ segments ending at some intermediate point $\tau_{N-1}$, plus the error of a single segment covering the remainder of the interval:
\begin{equation}
    E_N(b) = \min_{a \le \tau_{N-1} < b} \bigg[ E_{N-1}(\tau_{N-1}) + \mathcal{E}(\tau_{N-1}, b) \bigg],
\end{equation}
where $\mathcal{E}(u, v)$ calculates the minimum squared error of fitting a single line to $x(t)$ on the interval $[u, v]$.

This recursion reveals the key insight of dynamic programming: the globally optimal decision at the final step depends only on the optimal accumulated cost of the previous steps and the immediate cost of the current decision. By solving these subproblems sequentially, the procedure guarantees a globally optimal segmentation of the signal. The following section extends this framework to the problem of segmented linear least-squares estimation for beamforming.

\subsection{Segmented Linear Least Squares Estimation}

Bellman's optimality principle translates to the discrete-time problem of linear least-squares estimation. Consider a scenario where the system observes a target signal $d[t]$ and a corresponding regressor vector $\mathbf{x}[t] \in \mathbb{R}^p$ at each time step $t$. A standard linear estimator seeks a weight vector $\mathbf{w}$ that minimizes the squared error between the target and the linear estimate.

In a non-stationary environment, a single weight vector proves insufficient. Instead, the framework partitions the observation interval $[1, T]$ into a set of non-overlapping segments, applying a distinct, locally optimal weight vector within each. Let $e_{i,j}$ denote the minimum residual sum of squares (RSS) for a single linear estimator fit to the data over the interval $[i, j]$:
\begin{equation}
    e_{i,j} = \min_{\mathbf{w}} \sum_{k=i}^{j} \left| d[k] - \mathbf{w}^\top \mathbf{x}[k] \right|^2.
\end{equation}

If the number of segments remains unknown, simply minimizing the sum of these errors leads to overfitting, where the optimal solution trivializes to one segment per sample. To enforce parsimony, a penalty $C$ penalizes each segment. The objective seeks a partition set $\mathcal{P} = \{ [i_1, i_2], [i_2+1, i_3], \dots \}$ that minimizes the penalized cost:
\begin{equation}
    E = \min_{\mathcal{P}} \left( |\mathcal{P}| C + \sum_{[i_s, i_e] \in \mathcal{P}} e_{i_s, i_e} \right),
\end{equation}
where $|\mathcal{P}|$ defines the number of segments.

This global optimization problem solves recursively. Let $E[t]$ represent the minimum penalized cost for the data subsequence up to time $t$. The Bellman optimality equation is given by:
\begin{equation}
    E[t] = \min_{1 \le i \le t} \Big( e_{i, t} + C + E[i-1] \Big),
\end{equation}
with the base case $E[0] = 0$. This equation indicates that the optimal segmentation ending at time $t$ consists of a final segment starting at $i$, appended to the optimal segmentation of the data up to $i-1$.

Algorithm \ref{alg:SLS} details the batch implementation of this method. While computationally intensive due to the calculation of least-squares solutions for all candidate segments, it establishes the globally optimal baseline against which online approximations compare.

\begin{algorithm}[ht!]
\caption{Segmented Least Squares Estimation}\label{alg:SLS}
\begin{algorithmic}[1]
\Require $\mathbf{X} \in \mathbb{R}^{p \times T}$ (Regressor matrix), $\mathbf{d} \in \mathbb{R}^T$ (Target signal), $C$ (Penalty), $\delta$ (Regularization)
\Ensure $\hat{\mathbf{d}} \in \mathbb{R}^T$ (Estimate), $\mathcal{P}$ (Partitions)

\State \textbf{Initialize:}
\For{$j = 0$ \textbf{to} $T$}
    \State $E[j] \gets \infty$
\EndFor
\State $E[0] \gets 0$
\For{$j = 0$ \textbf{to} $T$}
    \State $P[j] \gets -1$
    \State $W[j] \gets \text{None}$
\EndFor

\For{$j = 1$ \textbf{to} $T$} \Comment{Iterate through all possible end points}
    \For{$i = 1$ \textbf{to} $j$} \Comment{Iterate through all possible start points}
        \State $\mathbf{X}_{\text{seg}} \gets \mathbf{X}[:, i:j]$ \Comment{Extract segment data}
        \State $\mathbf{d}_{\text{seg}} \gets \mathbf{d}[i:j]$
        
        \State $\mathbf{R} \gets \mathbf{X}_{\text{seg}} \mathbf{X}_{\text{seg}}^\top + \delta \mathbf{I}$ \Comment{Solve Ridge Regression}
        \State $\mathbf{r} \gets \mathbf{X}_{\text{seg}} \mathbf{d}_{\text{seg}}$
        \State $\mathbf{w} \gets \mathbf{R}^{-1} \mathbf{r}$
        
        \State $\hat{\mathbf{d}}_{\text{seg}} \gets \mathbf{w}^\top \mathbf{X}_{\text{seg}}$ \Comment{Compute segment error}
        \State $\text{error} \gets \|\mathbf{d}_{\text{seg}} - \hat{\mathbf{d}}_{\text{seg}}\|^2$
        
        \State $\text{cost} \gets \text{error} + C + E[i-1]$ \Comment{Bellman update}
        
        \If{$\text{cost} < E[j]$}
            \State $E[j] \gets \text{cost}$
            \State $P[j] \gets i-1$ \Comment{Store split point}
            \State $W[j] \gets \mathbf{w}$ \Comment{Store weights}
        \EndIf
    \EndFor
\EndFor

\State \textbf{Traceback:}
\State $\mathcal{P} \gets \emptyset$, $\text{weights} \gets \emptyset$
\State $j \gets T$
\While{$j > 0$}
    \State $i \gets P[j] + 1$
    \State Append $[i, j]$ to $\mathcal{P}$
    \State Append $W[j]$ to $\text{weights}$
    \State $j \gets P[j]$
\EndWhile
\State Reverse $\mathcal{P}$ and $\text{weights}$

\State \textbf{Compute Output:}
\State $\hat{\mathbf{d}} \gets \mathbf{0}$
\For{each $([i, j], \mathbf{w})$ in $(\mathcal{P}, \text{weights})$}
    \State $\hat{\mathbf{d}}[i:j] \gets \mathbf{w}^\top \mathbf{X}[:, i:j]$
\EndFor
\State \Return $\hat{\mathbf{d}}, \mathcal{P}$
\end{algorithmic}
\end{algorithm}

\subsection{Online Segmented Recursive Least Squares}

The Online Segmented Recursive Least Squares (OSRLS) algorithm \cite{osrls} provides a causal approximation to the batch SLS solution. Instead of optimizing the segmentation over the entire observation interval, OSRLS operates sequentially, making locally optimal decisions to detect change points in real-time.

The core assumption of OSRLS implies that the most recently detected change point, denoted $t_p$, acts as a correct anchor. This transforms the global optimization problem into a local search for the next optimal boundary. At each time step $t$, the algorithm evaluates whether to extend the current segment starting at $t_p$, or to introduce a new partition at some intermediate point $i \in [t_p, t]$.

The recursive cost function evaluated at each step formulates as:
\begin{equation}
    E(t) \approx \min_{t_p \le i \le t} \Big( \mathcal{E}(i, t) + C + \mathcal{E}(t_p, i-1) \Big) + E(t_p-1).
\end{equation}
Here, $E(t_p-1)$ represents the frozen cumulative cost up to the previous split point. This greedy strategy reduces computational complexity from $O(T^2)$ to $O(T)$ (or $O(1)$ with a fixed search window), functioning as an approximation of the global optimal structure. Decisions made in the past remain irrevocable, analogous to path pruning in sequential Viterbi decoding.

Algorithm \ref{alg:OSRLS} outlines the procedure. The algorithm maintains a set of parallel RLS filters, each hypothesizing a different start time for the current segment. At every time step, it updates these filters and compares their penalized costs. If a filter starting later than the current active partition achieves a lower cost (accounting for the penalty $C$), the algorithm declares a change point and switches to that filter.

\begin{algorithm}[ht!]
\caption{Online Segmented Recursive Least Squares (OSRLS)}\label{alg:OSRLS}
\begin{algorithmic}[1]
\Require $\mathbf{X} \in \mathbb{R}^{p \times T}$ (Regressor matrix), $\mathbf{d} \in \mathbb{R}^T$ (Target signal), $C$ (Penalty), $\delta$ (Loading), $\tau$ (Min segment length)
\Ensure $\hat{\mathbf{d}} \in \mathbb{R}^T$ (Output estimate)

\State \textbf{Initialize:}
\For{$i = 0$ \textbf{to} $T{-}1$}
    \State $\mathbf{w}[i] \gets \mathbf{0}_{p}$, $\mathbf{R}^{-1}[i] \gets \mathbf{I}_p / \delta$
    \State $\mathbf{r}[i] \gets \mathbf{0}_{p}$, $\rho[i] \gets 0$
\EndFor
\State $\text{cur} \gets 0$, $E[-1] \gets 0$, $\hat{\mathbf{d}} \gets \mathbf{0}_T$

\For{$n = 0$ \textbf{to} $T{-}1$}
    \State $\mathbf{u} \gets \mathbf{X}[:, n]$
    \State $\hat{\mathbf{d}}[n] \gets \mathbf{w}[\text{cur}]^\top \mathbf{u}$ \Comment{Predict using current active model}

    \State $E_{\min} \gets \infty$, $\text{best} \gets \text{cur}$

    \For{$i = \text{cur}$ \textbf{to} $n$} \Comment{Update candidate models}
        \State $\mathbf{k} \gets \mathbf{R}^{-1}[i] \mathbf{u} \,/\, (1 + \mathbf{u}^\top \mathbf{R}^{-1}[i] \mathbf{u})$ \Comment{RLS Update}
        \State $\mathbf{R}^{-1}[i] \gets \mathbf{R}^{-1}[i] - \mathbf{k} \mathbf{u}^\top \mathbf{R}^{-1}[i]$
        
        \State $\mathbf{r}[i] \gets \mathbf{r}[i] + \mathbf{d}[n] \mathbf{u}$
        \State $\rho[i] \gets \rho[i] + \mathbf{d}[n]^2$
        
        \State $\mathbf{w}[i] \gets \mathbf{R}^{-1}[i] \mathbf{r}[i]$ \Comment{Implicit weight update}

        \State $e_i \gets \rho[i] - \mathbf{r}[i]^\top \mathbf{w}[i]$ \Comment{Compute segment cost}
        \State $E_{\text{total}} \gets E[i{-}1] + C + e_i$ \Comment{Do not include C for current segment}

        \If{$E_{\text{total}} < E_{\min}$}
            \State $E_{\min} \gets E_{\text{total}}$
            \State $\text{best} \gets i$
        \EndIf
    \EndFor

    \State $E[n] \gets E_{\min}$

    \If{$(\text{best} - \text{cur}) > \tau$} \Comment{Check for change point detection}
        \State $\text{cur} \gets \text{best}$ \Comment{Switch to new optimal segment}
    \EndIf
\EndFor
\State \Return $\hat{\mathbf{d}}$
\end{algorithmic}
\end{algorithm}

\section{Segmented Beamforming}
\subsection{Optimality Equation}
The fundamental insight of the Capon beamformer relies on the minimization of output power to suppress interference. In practical settings where the system estimates the covariance matrix from data, the standard implementation effectively minimizes the sum of squared outputs over the entire observation window. For a given batch of data of length $T$, the Capon objective takes the form:
\begin{equation}
    J_1(T) = \min_{\mathbf{w}} \sum_{t=1}^T |\mathbf{w}^\top \mathbf{x}[t]|^2 \quad \text{such that} \quad \mathbf{w}^\top \boldsymbol{\nu} = 1.
\end{equation}
The linear distortionless constraint $\mathbf{w}^\top \boldsymbol{\nu} = 1$ applies implicitly to all subsequent minimization problems.

The segmented formulation posits that this standard setup represents a specific, restrictive case of a more general optimization problem: the temporally segmented minimum variance beamformer. By applying the principle of optimality, the Capon beamformer generalizes to piecewise-stationary environments.

Let $\mathcal{E}(i, j)$ denote the minimum output power achievable by a single constant weight vector over the interval $[i, j]$:
\begin{equation}
    \mathcal{E}(i, j) = \min_{\mathbf{w}} \sum_{t=i}^j |\mathbf{w}^\top \mathbf{x}[t]|^2.
\end{equation}
The standard Capon beamformer corresponds to the 1-segment solution, $J_1(T) = \mathcal{E}(1, T)$.

For a 2-segment partition, the objective seeks an intermediate boundary $\tau$ and two distinct weight vectors that minimize the total power:
\begin{equation}
    J_2(T) = \min_{1 \le \tau < T} \left[ \min_{\mathbf{w}_1} \sum_{t=1}^{\tau} |\mathbf{w}_1^\top \mathbf{x}[t]|^2 + \min_{\mathbf{w}_2} \sum_{t=\tau+1}^{T} |\mathbf{w}_2^\top \mathbf{x}[t]|^2 \right].
\end{equation}
Using the interval error notation, this simplifies to:
\begin{equation}
    J_2(T) = \min_{1 \le \tau < T} \big[ \mathcal{E}(1, \tau) + \mathcal{E}(\tau+1, T) \big].
\end{equation}

Generalizing to $N$ segments, the problem exhibits optimal substructure. The minimum output power for an $N$-segment partition, denoted $J_N(T)$, resolves via the Bellman optimality equation:
\begin{equation}
    J_N(T) = \min_{1 \le \tau < T} \Big[ J_{N-1}(\tau) + \mathcal{E}(\tau+1, T) \Big].
\end{equation}

This derivation establishes the formulation as a minimum-power solution for time-varying environments. It explicitly recognizes that the standard Capon beamformer acts as a stationary (1-segment) approximation of a dynamic acoustic scene. By relaxing the constraint that a single weight vector must apply across all snapshots, the Batch Segmented Beamformer (BSB) generalizes the MVDR principle, allowing the filter to adapt to discrete changes in the interference structure.

\subsection{Penalized Segmented Beamforming}

The previous subsection assumed the number of stationary segments $N$ remained known a priori. In practice, both the number and the temporal locations of these stationary regions must infer directly from the data. If the system minimizes the total output power without constraint on $N$, the trivial solution selects a distinct beamformer for every single snapshot ($N=T$), reducing the error to zero but failing to suppress interference due to the lack of covariance averaging.

To avoid this degenerate overfitting, the framework introduces a regularization term $C$ that penalizes model complexity. This term imposes a fixed cost for each new segment, enforcing a tradeoff between minimizing output power (fitting the data) and maximizing segment length (covariance averaging).

Formally, the penalized segmented beamforming objective seeks a partition set $\mathcal{P}$ that minimizes:
\begin{equation}
    E = \min_{\mathcal{P}} \left( |\mathcal{P}| C + \sum_{[i_s, i_e] \in \mathcal{P}} \mathcal{E}(i_s, i_e) \right),
\end{equation}
where $|\mathcal{P}|$ acts as the number of segments, $C$ functions as the penalty parameter, and $\mathcal{E}(i_s, i_e)$ defines the minimum output power achievable by a Capon beamformer over the interval $[i_s, i_e]$. A smaller $C$ encourages frequent segmentation to track rapid changes, while a larger $C$ favors longer, stable segments.

This global optimization problem exhibits optimal substructure and solves via dynamic programming. The minimum penalized cost $E[t]$ for the data up to time $t$ satisfies the Bellman optimality equation:
\begin{equation}
    E[t] = \min_{1 \le i \le t} \Big( \mathcal{E}(i, t) + C + E[i-1] \Big).
\end{equation}
Here, $\mathcal{E}(i, t)$ represents the output power of the optimal batch MVDR beamformer computed on snapshots $\mathbf{x}[i], \dots, \mathbf{x}[t]$. This recursion guarantees finding the globally optimal piecewise-stationary solution, balancing interference suppression against estimation variance. Algorithm \ref{alg:SLS_BF_efficient} details the execution.

\begin{algorithm}[ht!]
\caption{Batch Segmented Beamformer (BSB)}\label{alg:SLS_BF_efficient}
\begin{algorithmic}[1]
\Require $\mathbf{X} \in \mathbb{R}^{p \times T}$ (Snapshots), $\boldsymbol{\nu} \in \mathbb{R}^{p}$ (Steering vector), $C$ (Penalty), $\delta$ (Diagonal loading)
\Ensure $\mathbf{z} \in \mathbb{R}^{T}$ (Beamformed output), $\mathcal{P}$ (Partitions)

\State \textbf{Initialize:}
\For{$j = 0$ \textbf{to} $T$}
    \State $E[j] \gets \infty$
\EndFor
\State $E[0] \gets 0$
\For{$j = 0$ \textbf{to} $T$}
    \State $P[j] \gets -1$
    \State $W[j] \gets \text{None}$
\EndFor

\For{$j = 1$ \textbf{to} $T$} \Comment{Iterate through all possible segment end points}
  \For{$i = 1$ \textbf{to} $j$} \Comment{Iterate through all possible segment start points}
    \State $\mathbf{X}_{\text{seg}} \gets \mathbf{X}[:, i:j]$ \Comment{Extract segment data}
    
    \State $\mathbf{R} \gets \mathbf{X}_{\text{seg}} \mathbf{X}_{\text{seg}}^\top + \delta \mathbf{I}$ \Comment{Compute MVDR weights}
    \State $\mathbf{u} \gets \mathbf{R}^{-1}\boldsymbol{\nu}$
    \State $\mathbf{w} \gets \mathbf{u} / (\boldsymbol{\nu}^\top \mathbf{u})$
    
    \State $\mathbf{z}_{\text{seg}} \gets \mathbf{w}^\top \mathbf{X}_{\text{seg}}$ \Comment{Compute output power (cost)}
    \State $\text{error} \gets \|\mathbf{z}_{\text{seg}}\|^2$
    
    \State $\text{cost} \gets \text{error} + C + E[i-1]$
    
    \If{$\text{cost} < E[j]$}
      \State $E[j] \gets \text{cost}$
      \State $P[j] \gets i-1$ \Comment{Store optimal split point}
      \State $W[j] \gets \mathbf{w}$ \Comment{Store optimal weights}
    \EndIf
  \EndFor
\EndFor

\State \textbf{Traceback:}
\State $\mathcal{P} \gets \emptyset$, $\text{weights} \gets \emptyset$
\State $j \gets T$
\While{$j > 0$}
  \State $i \gets P[j] + 1$
  \State Append $[i, j]$ to $\mathcal{P}$
  \State Append $W[j]$ to $\text{weights}$
  \State $j \gets P[j]$
\EndWhile
\State Reverse $\mathcal{P}$ and $\text{weights}$

\State \textbf{Synthesize output:}
\State $\mathbf{z} \gets \mathbf{0} \in \mathbb{R}^{T}$
\For{each $([i, j], \mathbf{w})$ in $(\mathcal{P}, \text{weights})$}
  \State $\mathbf{z}[i:j] \gets \mathbf{w}^\top \mathbf{X}[:, i:j]$
\EndFor

\State \Return $\mathbf{z}, \mathcal{P}$
\end{algorithmic}
\end{algorithm}

\subsection{Online Segmented Beamformer}

While the penalized segmented beamformer achieves a globally optimal segmentation in the batch setting, solving the full Bellman recursion requires $O(T^2)$ operations, which proves infeasible for real-time applications. To enable sequential processing, the Online Segmented Beamformer (OSB) adopts a causal approximation analogous to the OSRLS method.

The OSB maintains the most recently detected segmentation point, denoted $t_p$, and establishes it as a fixed anchor rather than re-evaluating the entire history of boundaries. At each new snapshot, the algorithm compares the cost of extending the current segment against the cost of initiating a new segment at a later point. The approximated recursive cost function equals:
\begin{equation}
    E(t) \approx \min_{t_p \le i \le t} \Big( \mathcal{E}(i, t) + C + \mathcal{E}(t_p, i-1) \Big) + E(t_p-1).
\end{equation}
This greedy approximation reduces the complexity from quadratic to linear in time $t$ (or constant if constrained by a fixed search window). Solving this approximation gap entirely requires the batch dynamic programming pass over the full observation record, which forfeits causal operation. The theoretical bounds in Section \ref{sec:theory} address the worst-case divergence between this online approximation and the exact batch solution.

Operationally, the algorithm maintains a set of candidate MVDR beamformers running in parallel, each hypothesizing a different start time for the current segment. The inverse covariance matrix $\mathbf{S}^{-1}$ and weight vector $\mathbf{w}$ for each candidate update recursively using the Woodbury identity. Whenever the accumulated cost of a candidate starting at a later time (plus the penalty $C$) drops below the cost of the current active segment, the algorithm declares a change point. The beamformer then switches to the new optimal segment, effectively resetting the covariance estimate to adapt to the new acoustic regime. Algorithm \ref{alg:OSB} provides the execution logic.

\begin{algorithm}[ht!]
\caption{Online Segmented Beamformer (OSB)}\label{alg:OSB}
\begin{algorithmic}[1]
\Require $\mathbf{X} \in \mathbb{R}^{p \times T}$ (Snapshots), $\boldsymbol{\nu} \in \mathbb{R}^{p}$ (Steering), $C$ (Penalty), $\delta$ (Loading), $\tau$ (Min segment length)
\Ensure $\mathbf{z} \in \mathbb{R}^{T}$ (Output), $\mathcal{I}$ (Partition indices)

\State \textbf{Initialize per-start states:}
\For{$i = 0$ \textbf{to} $T{-}1$}
    \State $\mathbf{S}^{-1}[i] \gets \mathbf{I}_p / \delta$
    \State $\mathbf{u}[i] \gets \boldsymbol{\nu} / \delta$ \Comment{Numerator state: $\mathbf{S}^{-1}\boldsymbol{\nu}$}
    \State $\rho[i] \gets \boldsymbol{\nu}^\top \mathbf{u}[i]$ \Comment{Denominator state: $\boldsymbol{\nu}^\top \mathbf{S}^{-1}\boldsymbol{\nu}$}
    \State $\mathbf{w}[i] \gets \mathbf{u}[i] / \rho[i]$
    \State $J[i] \gets 0$ \Comment{Accumulated segment cost (output power)}
\EndFor
\State $\mathcal{I} \gets [0]$, $\text{cur} \gets 0$
\State $E[-1] \gets 0$, $\mathbf{z} \gets \mathbf{0}_T$

\For{$n = 0$ \textbf{to} $T{-}1$}
  \State $\mathbf{x} \gets \mathbf{X}[:, n]$
  \State $\mathbf{z}[n] \gets \mathbf{w}[\text{cur}]^\top \mathbf{x}$ \Comment{Filter output using active model}

  \State $E_{\min} \gets \infty$, $\text{best} \gets \text{cur}$

  \For{$i = \text{cur}$ \textbf{to} $n$} \Comment{Update candidates}
    \State $\mathbf{k} \gets \mathbf{S}^{-1}[i] \mathbf{x} \,/\, (1 + \mathbf{x}^\top \mathbf{S}^{-1}[i] \mathbf{x})$ \Comment{Woodbury update}
    \State $\mathbf{U} \gets \mathbf{k} \mathbf{x}^\top \mathbf{S}^{-1}[i]$
    \State $\mathbf{S}^{-1}[i] \gets \mathbf{S}^{-1}[i] - \mathbf{U}$
    
    \State $\mathbf{u}[i] \gets \mathbf{u}[i] - \mathbf{U} \boldsymbol{\nu}$ \Comment{Update MVDR weights}
    \State $\rho[i] \gets \boldsymbol{\nu}^\top \mathbf{u}[i]$
    \State $\mathbf{w}[i] \gets \mathbf{u}[i] / \rho[i]$
    
    \State $y \gets \mathbf{w}[i]^\top \mathbf{x}$ \Comment{Update cost}
    \State $J[i] \gets J[i] + y^2$

    \State $E_{\text{total}} \gets E[i{-}1] + C + J[i]$ \Comment{Do not include C for current segment}
    \If{$E_{\text{total}} < E_{\min}$}
      \State $E_{\min} \gets E_{\text{total}}$
      \State $\text{best} \gets i$
    \EndIf
  \EndFor

  \State $E[n] \gets E_{\min}$

  \If{$(\text{best} - \text{cur}) > \tau$}
    \State $\text{cur} \gets \text{best}$ \Comment{Switch to new optimal segment}
    \State Append $\text{best}$ to $\mathcal{I}$
  \EndIf
\EndFor

\State \Return $\mathbf{z}, \mathcal{I}$
\end{algorithmic}
\end{algorithm}

\section{Theoretical Analysis}
\label{sec:theory}

In this section, the analysis derives a formal regret bound for the OSB relative to the optimal BSB. While the batch formulation guarantees global optimality via dynamic programming, it remains inherently non-causal. The online algorithm operates greedily, making irrevocable segmentation decisions based on past data.

To quantify the performance penalty incurred by this causal restriction, the analysis employs the framework of \textit{Competitive Analysis} for individual sequences \cite{merhav2002limiting, switching, cesa2006prediction}. Unlike statistical analyses that rely on stationarity or ergodic assumptions, which fail in transient environments, this approach treats the sequence of sensor snapshots as deterministic. The cumulative loss of the online algorithm faces bounding relative to the best possible piecewise-stationary solution chosen in hindsight by an "Oracle" with full knowledge of the future.

\subsection{Problem Setup}

Let the sequence of received sensor snapshots conform to $\mathbf{x}_{1:T} = \{\mathbf{x}[1], \dots, \mathbf{x}[T]\}$, where $\mathbf{x}[t] \in \mathbb{R}^{d}$ and $d = p$ represents the beamformer degrees of freedom.

\subsubsection{Loss Functions: Sequential vs. Batch}
The instantaneous performance of a beamformer with weight vector $\mathbf{w}$ at time $t$ reflects the squared output power:
\begin{equation}
    \ell(\mathbf{w}, \mathbf{x}[t]) = |\mathbf{w}^\top \mathbf{x}[t]|^2.
\end{equation}

The analysis distinguishes between the \textit{sequential} error incurred by the online algorithm and the \textit{batch} error achieved by the Oracle.

\textbf{1. Sequential Prediction Error ($e_{a,b}$):}
For the online algorithm and the reference strategy, weights adapt causally (e.g., via RLS). The cumulative sequential error over an interval $[a, b]$ sums the instantaneous losses using the \textit{current} estimate $\hat{\mathbf{w}}_{t-1}$:
\begin{equation}
    e_{a,b} = \sum_{t=a}^{b} \ell(\hat{\mathbf{w}}_{t-1}, \mathbf{x}[t]).
\end{equation}
By convention, if $a > b$, $e_{a,b} = 0$.

\textbf{2. Minimum Batch Error ($E^*_{a,b}$):}
The Oracle selects the single best weight vector for the entire interval in hindsight. The minimum batch error formulates as:
\begin{equation}
    E^*_{a,b} = \min_{\mathbf{w}} \sum_{t=a}^{b} \ell(\mathbf{w}, \mathbf{x}[t]).
\end{equation}
Note that $E^*_{a,b} \le e_{a,b}$ always holds, as the sequential predictor suffers from parameter convergence costs.

\subsubsection{The Oracle (Batch) Objective}
Let $\mathcal{P}^*$ denote an arbitrary partition of the time horizon $[1, T]$ into $K^*$ segments, defined by change points $0 = \tau_0 < \tau_1 < \dots < \tau_{K^*} = T$. The \textit{Batch Penalized Cost} minimizes the sum of the batch errors plus a regularization penalty $C$:
\begin{equation}
    L_{Batch}(\mathcal{P}^*) = \sum_{j=1}^{K^*} \bigg( C + E^*_{\tau_{j-1}+1, \tau_j} \bigg).
\end{equation}

\subsubsection{The Greedy Potential Function}
The OSB minimizes a recursive cost potential $E(t)$. Let $t_p$ denote the \textbf{start index of the current active segment} (the index immediately following the last finalized cut).

At each time step $t$, the algorithm updates its potential by searching for the optimal start point $i$ for the current pending segment within the active window:
\begin{equation}
    \begin{split}
        E(t) = \min_{t_p \le i \le t} \Big( e_{i,t} + C + e_{t_p, i-1} \Big) + E(t_p-1),
    \end{split}
    \label{eq:greedy_potential}
\end{equation}
where $E(t_p-1)$ represents the frozen cumulative cost of all previously finalized segments. The realized cumulative loss of the algorithm matches exactly the final value of this potential: $L_{Alg}(T) = E(T)$.

\subsection{Regret Bound Derivation}

Theorem 1 proves that the greedy minimization inherent in Eq. \eqref{eq:greedy_potential} guarantees the algorithm performs competitively against the Oracle.

\textbf{Theorem 1 (Universal Regret Bound).} 
\textit{The cumulative regret of the OSB relative to the optimal BSB with $K^*$ segments bounds to:}
\begin{equation}
    \begin{split}
        L_{Alg}(T) - L_{Batch}(\mathcal{P}^*) \le \\
        K^* \left( \max(C, \tau L_{max}) + \frac{d}{2} \ln\left(\frac{T}{K^*}\right) + \Gamma \right),
    \end{split}
\end{equation}
\textit{where $\tau$ defines the stability constraint (minimum segment length), and $\Gamma$ acts as a constant related to initialization.}

\textbf{Proof.}
The proof constructs a \textit{Reference Strategy} $\pi_{ref}$ that mimics the Oracle partition $\mathcal{P}^*$ subject to the causality constraints of the online setting. Let $E_{Ref}(t)$ denote the cumulative cost of this strategy. The proof proceeds by strong induction on the Oracle change points $\{\tau_k\}$. It establishes that at every such boundary, the greedy algorithm's accumulated cost does not exceed the Reference Strategy's cost:
\begin{equation}
    E(\tau_k) \le E_{Ref}(\tau_k) \quad \forall k=1 \dots K^*.
\end{equation}

\subsubsection{Base Case}
At $t=0$, $E(0) = E_{Ref}(0) = 0$. The hypothesis holds trivially.

\subsubsection{Inductive Step: The General Case}
Assume that at the previous Oracle change point $\tau_{k-1}$, the online algorithm's accumulated cost satisfied $E(\tau_{k-1}) \le E_{Ref}(\tau_{k-1})$.

The analysis tracks cumulative cost over the interval $(\tau_{k-1}, \tau_k]$. Let $H(t)$ denote the cost trajectory of the Reference Strategy, which extends a single segment from $\tau_{k-1}$ to $t$ (imitating the Oracle). Its cost at the end of the interval takes the form:
\begin{equation}
    H(\tau_k) = E(\tau_{k-1}) + C + e_{\tau_{k-1}+1, \tau_k}.
\end{equation}
By the inductive hypothesis, $H(\tau_k) \le E_{Ref}(\tau_{k-1}) + C + e_{\tau_{k-1}+1, \tau_k} = E_{Ref}(\tau_k)$. The proof must establish $E(\tau_k) \le H(\tau_k)$.

Let $t_{p}$ designate the start index of the algorithm's \textit{current active window} at time $\tau_k$. Two scenarios govern the algorithm's behavior in this interval:

\textbf{Scenario 1: No Intermediate Cuts ($t_p \le \tau_{k-1} + 1$).}
The algorithm has not finalized any segments since before $\tau_{k-1}$. In this case, the minimization window $[t_p, \tau_k]$ includes the Oracle's cut point. Specifically, $i^* = \tau_{k-1}+1$ represents a valid candidate. Evaluating the potential function for this candidate yields:
\begin{align}
    E(\tau_k) &\le \text{Cost}(i*) \nonumber \\
              &= e_{\tau_{k-1}+1, \tau_k} + C + \underbrace{(e_{t_p, \tau_{k-1}} + E(t_p-1))}_{= E(\tau_{k-1})} \nonumber \\
              &= H(\tau_k) \le E_{Ref}(\tau_k).
\end{align}
Because the algorithm selects the global minimum, its cost cannot exceed the cost of this specific candidate path. Thus, the bound holds directly.

\textbf{Scenario 2: Over-Partitioning ($t_p > \tau_{k-1} + 1$).}
Suppose the algorithm finalized an intermediate segment (false alarm) at time $u \in (\tau_{k-1}, \tau_k)$. This decision implies that at time $u$, the greedy minimization found a split point that yielded a \textit{strictly lower} cost than extending the Reference path.

Mathematically, if the algorithm chose to split at $u$, it must follow that:
\begin{equation}
    E(u) < H(u).
\end{equation}
If $H(u)$ were lower, the greedy algorithm would have chosen to extend the segment corresponding to $H(u)$ (Scenario 1) rather than finalize a split. This establishes a new inequality anchor at $t=u$: $E(u) < H(u)$.

Now, consider the final step at $\tau_k$. The Reference Strategy remains constrained to continue its original single-segment path $H(\tau_k)$. The algorithm, however, optimizes from its new, lower baseline $E(u)$. The algorithm's extension from $u$ compares against the Reference path:
\begin{align}
    E(\tau_k) &\le E(u) + e_{u+1, \tau_k} \nonumber \\
              &< H(u) + e_{u+1, \tau_k} \nonumber \\
              &= (E(\tau_{k-1}) + C + e_{\tau_{k-1}+1, u}) + e_{u+1, \tau_k}.
\end{align}
By the super-additivity property of least squares errors, splitting a segment generally reduces total error compared to a single fit: $e_{\tau_{k-1}+1, \tau_k} \ge e_{\tau_{k-1}+1, u} + e_{u+1, \tau_k}$. Therefore:
\begin{equation}
    \begin{split}
        H(u) + e_{u+1, \tau_k} \le E(\tau_{k-1}) + C + e_{\tau_{k-1}+1, \tau_k} \\
        = H(\tau_k).
    \end{split}
\end{equation}
Thus, $E(\tau_k) < H(\tau_k) \le E_{Ref}(\tau_k)$. The algorithm's decision to over-partition only tightens the bound.

\subsubsection{Cost Analysis of the Reference Strategy}

Having established that $L_{Alg}(T) \le E_{Ref}(T)$, the analysis upper-bounds the cost of the Reference Strategy itself. $E_{Ref}(T)$ reflects the sum of costs for $K^*$ segments. For each segment $S_k^*$ spanning $[\tau_{k-1}+1, \tau_k]$, the cost comprises three parts: the batch error, the structural regret, and the parameter regret.

\textbf{1. Batch Error:}
This forms the baseline cost paid by the Oracle:
\begin{equation}
    \text{Baseline}_k = E^*_{\tau_{k-1}+1, \tau_k} + C.
\end{equation}

\textbf{2. Structural Regret (Delay and Penalty):}
At the transition $\tau_{k-1}$, the Reference Strategy must detect the change.
\begin{itemize}
    \item If the previous segment was long ($> \tau$), the strategy resets immediately, paying the standard penalty $C$.
    \item If the previous segment was short ($< \tau$), the stability constraint forces the strategy to wait. It uses the incorrect weight vector for $\tau$ samples, paying at most $\tau L_{max}$ in mismatch loss before resetting.
\end{itemize}
Thus, the transition cost bounds to:
\begin{equation}
    \mathcal{R}_{struct, k} \le \max(C, \tau L_{max}).
\end{equation}

\textbf{3. Parameter Regret (Learning Cost):}
Once the Reference Strategy resets, it executes an RLS filter of length $n_k = \tau_k - \tau_{k-1}$. The difference between the cumulative sequential error $e$ and the optimal batch error $E^*$ bounds according to Rissanen's stochastic complexity \cite{rissanen1986stochastic}. For a $d$-dimensional parameter vector:
\begin{equation}
    e_{\tau_{k-1}+1, \tau_k} - E^*_{\tau_{k-1}+1, \tau_k} \le \frac{d}{2} \ln(n_k) + \Gamma,
\end{equation}
where $\Gamma$ accounts for initialization regularization.

\subsubsection{Synthesis}
Summing the costs over all $K^*$ segments gives:
\begin{equation}
    \begin{split}
        L_{Alg}(T) \le \sum_{k=1}^{K^*} \Big( (E^*_k + C) + \mathcal{R}_{struct, k} + \mathcal{R}_{param, k} \Big) \\
        \le L_{Batch}(\mathcal{P}^*) + \sum_{k=1}^{K^*} \bigg( \max(C, \tau L_{max}) \\
        + \frac{d}{2} \ln(n_k) + \Gamma \bigg).
    \end{split}
\end{equation}
To obtain the worst-case bound, the formulation maximizes the term $\sum \ln(n_k)$ subject to the constraint $\sum n_k = T$. By Jensen's inequality, the sum maximizes when all segments have equal length $n_k = T/K^*$. Rearranging terms yields the theorem. \hfill $\square$

\section{Simulations}
To empirically validate the theoretical bounds established in Section \ref{sec:theory} and to evaluate the practical efficacy of the Segmented Beamforming framework, the evaluation utilizes a comprehensive suite of numerical simulations. 

The core vulnerability of traditional adaptive beamforming, specifically the reliance on fixed-memory integration windows, becomes exposed in environments where the rate, duration, and spatial distribution of interference change unpredictably. To investigate this phenomenon, the simulations systematically isolate and test the algorithms against distinct modalities of non-stationarity. The experimental progression structures as follows:

\begin{enumerate}
    \item \textbf{Batch Formulation Validation (Section VII-A):} The evaluation verifies the fundamental hypothesis in a controlled, abrupt-change environment. By comparing the BSB against a genie-aided Segmented Least Squares (SLS) predictor, this scenario confirms that output power minimization serves as an effective proxy for estimation error, ensuring the global objective function correctly identifies true boundaries of stationarity.
    
    \item \textbf{Demonstrative Example of Online Segmented Beamforming (Section VII-B):} This evaluation aims to verify the online segmented beamforming hypothesis. We generate an abruptly changing scenario with beampatterns demonstrating the OSB's ability to place nulls. 
    
    \item \textbf{Piecewise Constant Bearing (Section VII-C):} To evaluate spatial transience, this scenario isolates the effect of interferers abruptly changing locations while maintaining a constant block duration. It demonstrates how standard fixed-window beamformers fail during sudden spatial shifts, whereas the OSB algorithm resets to prevent the smearing of spatial nulls.
    
    \item \textbf{Piecewise Constant Time (Section VII-D):} To challenge the assumption of a constant stationarity timescale, this scenario subjects the algorithms to active interference blocks of randomized, unpredictable durations. This tests the framework's ability to dynamically scale its effective memory length, proving it reacts during short bursts and stabilizes during long periods of stationarity.
    
    \item \textbf{Stochastic Birth-Death Process (Section VII-E):} Synthesizing spatial and temporal transience, this environment models a dynamic, Markov-driven interference field. This demonstrates the real-time regret-minimizing behavior of the OSB, showing its capacity to rapidly track continuous state changes and out-compete fixed-memory baselines.
\end{enumerate}

Collectively, this sequence of evaluations demonstrates that the Segmented Beamformer operates as a universal spatial filter. By continuously evaluating local stationarity and greedily managing the bias-variance tradeoff directly from the data, the framework consistently matches or exceeds the performance of hindsight-optimized fixed sliding windows.

\subsection{Batch Performance of Segmented Beamforming}

To verify that output power minimization serves as a valid proxy for true estimation error, this first simulation evaluates the BSB in an abrupt-change environment. A Uniform Linear Array (ULA) with 9 elements operates with an inter-element spacing of $d = 0.2$ m, corresponding to a design frequency of $3600$ Hz given a speed of sound $c = 1440$ m/s. The sampling rate equals 200 Hz.

The acoustic scene consists of a continuous desired target signal arriving from broadside (directional cosine $\cos \theta = 0$) and multiple interferers that enter and exit the scene in bursts. The interference state changes every 150 snapshots. In each Monte Carlo trial, the locations of the interferers resolve uniformly at random, and their Signal-to-Noise Ratios (SNR) draw uniformly from $20$ to $25$ dB. The target SNR remains fixed at $-9$ dB.

The BSB compares against the standard sample-matrix inversion (SMI) Capon beamformer, which utilizes a single covariance estimate averaged over the entire observation window. Comparisons also include a genie-aided Segmented Least Squares (SLS) predictor that accesses the desired signal waveform (serving as an upper bound on performance), whereas the Segmented Beamformer relies solely on the steering vector constraint.

\begin{figure}[htbp]
    \centering
    \includegraphics[width=\linewidth]{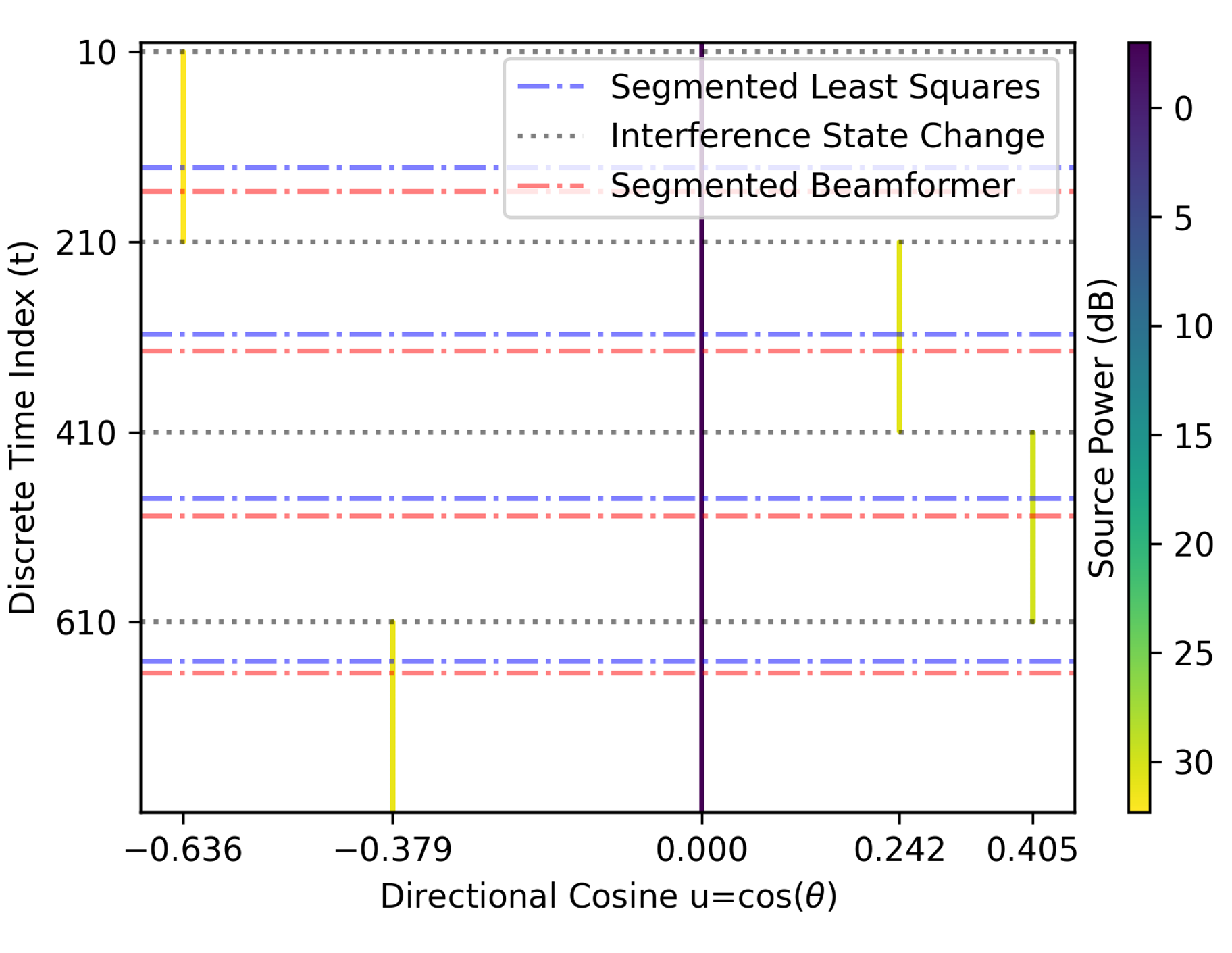}
    \caption{Bearing-time record of the simulated environment showing abrupt interference transitions. The vertical lines indicate the segmentation boundaries detected by the BSB, demonstrating its ability to recover the temporal structure of the interference.}
    \label{fig:SLS_Partitions}
\end{figure}

Figure \ref{fig:SLS_Partitions} visualizes the bearing-time record for a representative trial. The figure overlays the ground truth interference locations with the segmentation boundaries detected by the algorithm. The Segmented Beamformer successfully identifies the piecewise-stationary regions of the received signal. The detected boundaries align closely with those found by the optimal SLS predictor, confirming that the output power minimization objective effectively proxies for estimation error minimization.

The evaluation quantifies performance using the Mean Squared Error (MSE) of the signal estimate and the cumulative output power. The BSB achieves an MSE reduction of approximately $3$ dB compared to the standard Capon beamformer. This improvement stems from the segmented approach's ability to discard covariance data from inactive interference periods, preventing null placement in directions that no longer jam the array.

\begin{figure}[htbp]
    \centering
    \includegraphics[width=\linewidth]{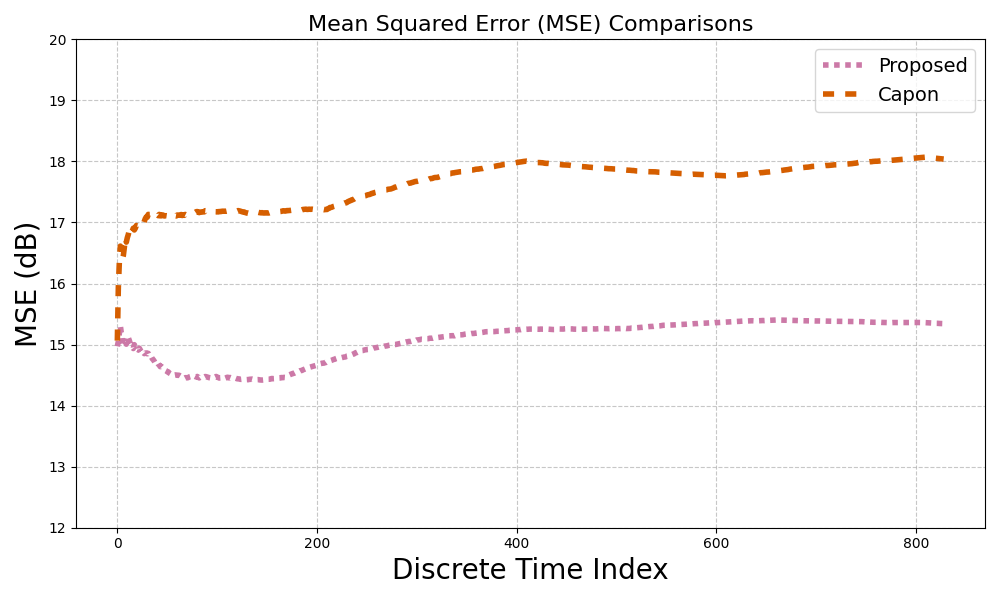}
    \caption{Average Mean Squared Error (MSE) across 100 Monte Carlo trials. The performance gap highlights the consistency of the segmented approach. A lower mean-squared error corresponds with better performance.}
    \label{fig:SLS_MSE_MC}
\end{figure}

\begin{figure}[htbp]
    \centering
    \includegraphics[width=\linewidth]{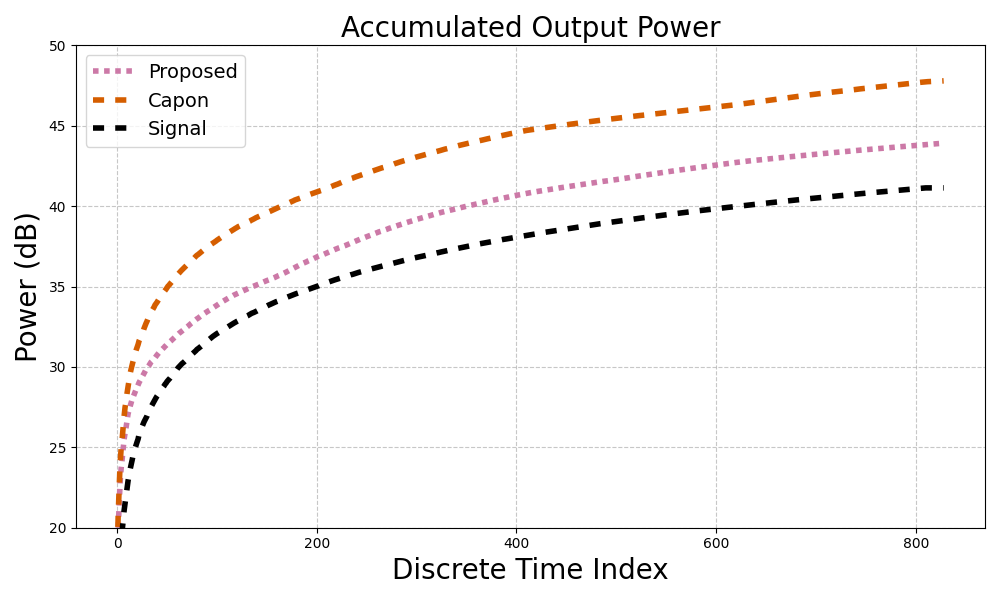}
    \caption{Average accumulated output power across 100 Monte Carlo trials. The batch Segmented beamformer is accumulates less power than the batch Capon beamformer, which indicates better performance under a unity gain constraint. }
    \label{fig:SLS_power_MC}
\end{figure}

To validate robustness, 100 Monte Carlo trials execute. Figure \ref{fig:SLS_MSE_MC} and Figure \ref{fig:SLS_power_MC} present the ensemble average results. The BSB consistently tracks the lower bound of the output power, adapting the integration time to match the stationarity of the environment.

\subsection{Demonstrative Evaluation of Online Segmented Beamforming}

To empirically validate the transition behavior and nulling capabilities of the Online Segmented Beamformer (OSB), this section analyzes the algorithm in a controlled scenario featuring distinct spatial shifts. Fixed-window algorithms natively struggle with abrupt environmental changes, necessitating an evaluation of how dynamic segmentation identifies and reacts to these transitions. Tracking the internal anchor state of the algorithm provides direct insight into how the sequential mechanism manages the covariance estimation window during sudden interference changes.

The online formulation greedily approximates the global Bellman optimality equation by fixing the anchor point of the currently active segment. This causal restriction inherently carries an approximation gap compared to the batch solution, as the online algorithm cannot irrevocably revise past segmentation boundaries. Solving this approximation gap entirely requires a non-causal batch process over the full observation record.

\begin{figure}[htbp]
    \centering
    \includegraphics[width=\linewidth]{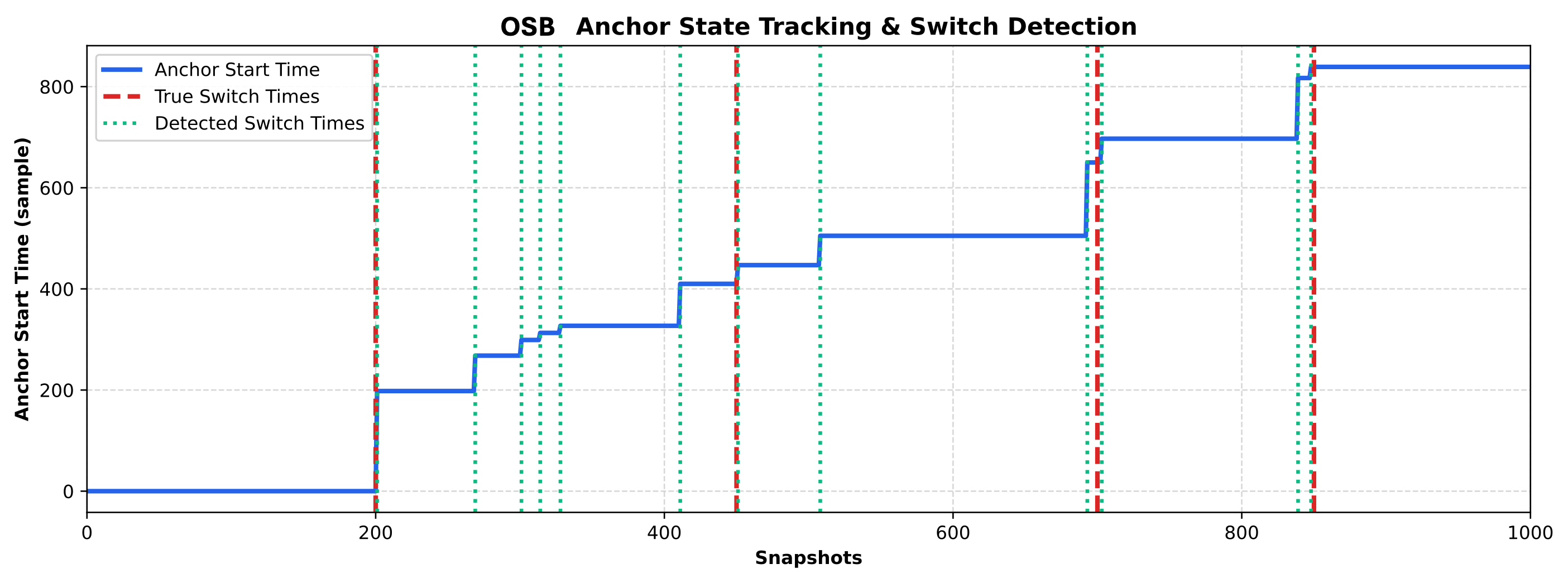}
    \caption{Demonstration of the changepoint detection of the Online Segmented Beamformer (OSB). The visualization compares detected changepoints against the ground truth. Note because OSB is a greedy algorithm it lacks the abilitiy to correct its decisions in hindsight. However, the algorithm correctly resets the covariance matrix at the true changepoints, as well. }
    \label{fig:changepoint_demonstration}
\end{figure}

\begin{figure}[htbp]
    \centering
    \includegraphics[width=\linewidth]{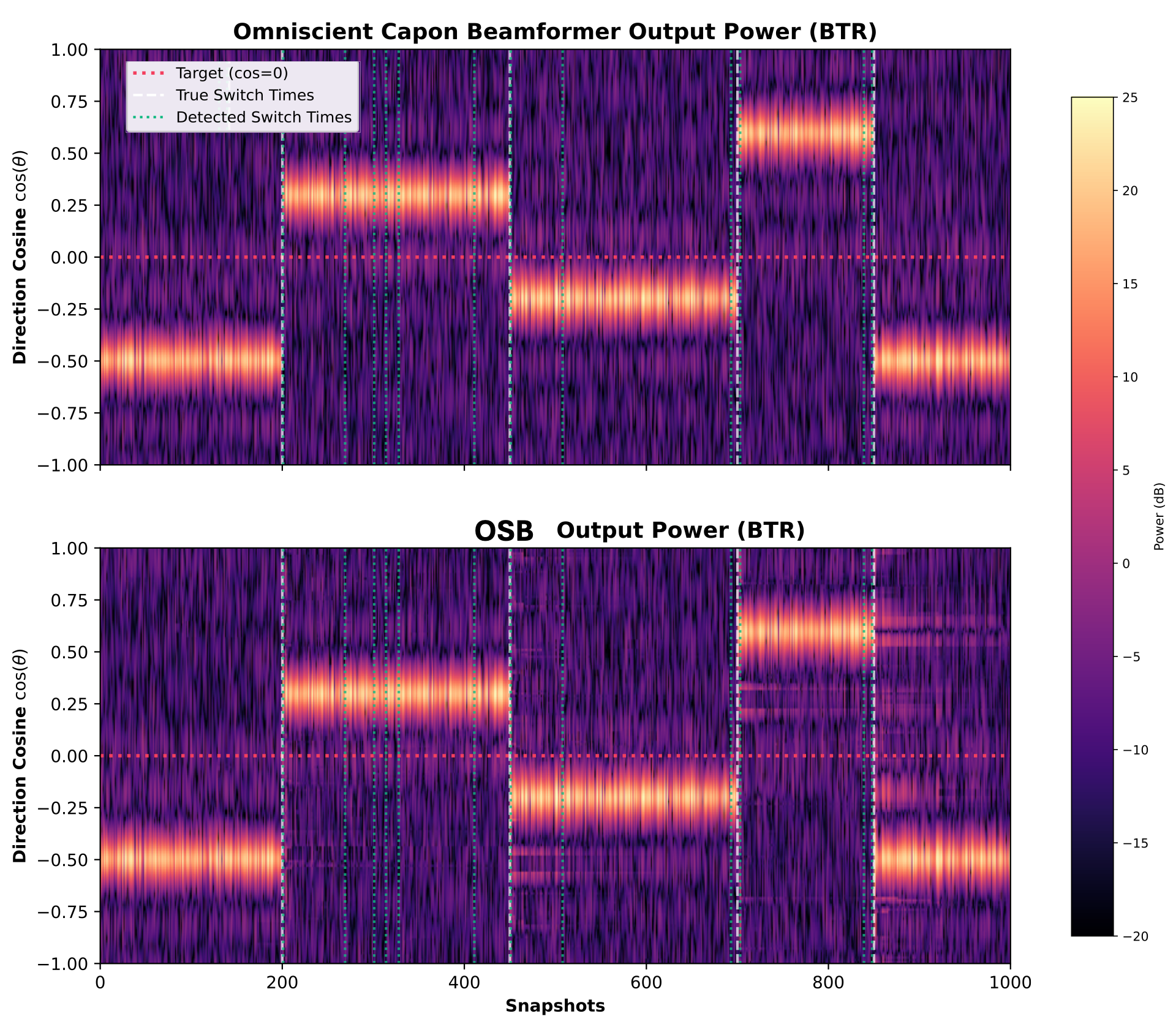}
    \caption{Demonstrative evaluation of the Online Segmented Beamformer (OSB). The visualization compares the spatial-temporal BTR estimates against the ground truth.}
    \label{fig:demonstrative_example}
\end{figure}

The simulation establishes an environment with a target signal arriving from broadside, corresponding to a direction cosine of $\cos(\theta) = 0$. Transient interferers shift spatial locations at discrete true switch times, specifically at snapshots $t \in \{200, 450, 700, 850\}$. The OSB monitors the acoustic scene and updates its anchor start time, representing the initialization point of the current sample covariance matrix. A discrete jump in this anchor state signifies the detection of a statistical change point, prompting the algorithm to flush stale data and begin a new integration window. Figure \ref{fig:changepoint_demonstration} illustrates the comprehensive behavior of the OSB during this sequence. Figure \ref{fig:demonstrative_example} contrasts the estimated Bearing Time Record (BTR) of the OSB against an omniscient Capon beamformer. The omniscient bound utilizes the exact instantaneous ensemble covariance matrix, serving as the theoretical performance ceiling.

The OSB identifies the true switch times, evident by the alignment between the true transition markers and the algorithm's detected boundaries in the tracked anchor state. The resulting BTR closely mirrors the omniscient Capon output, confirming that the algorithm avoids the temporal smearing characteristic of fixed-window estimators.

\begin{figure}[htbp]
    \centering
    \includegraphics[width=\linewidth]{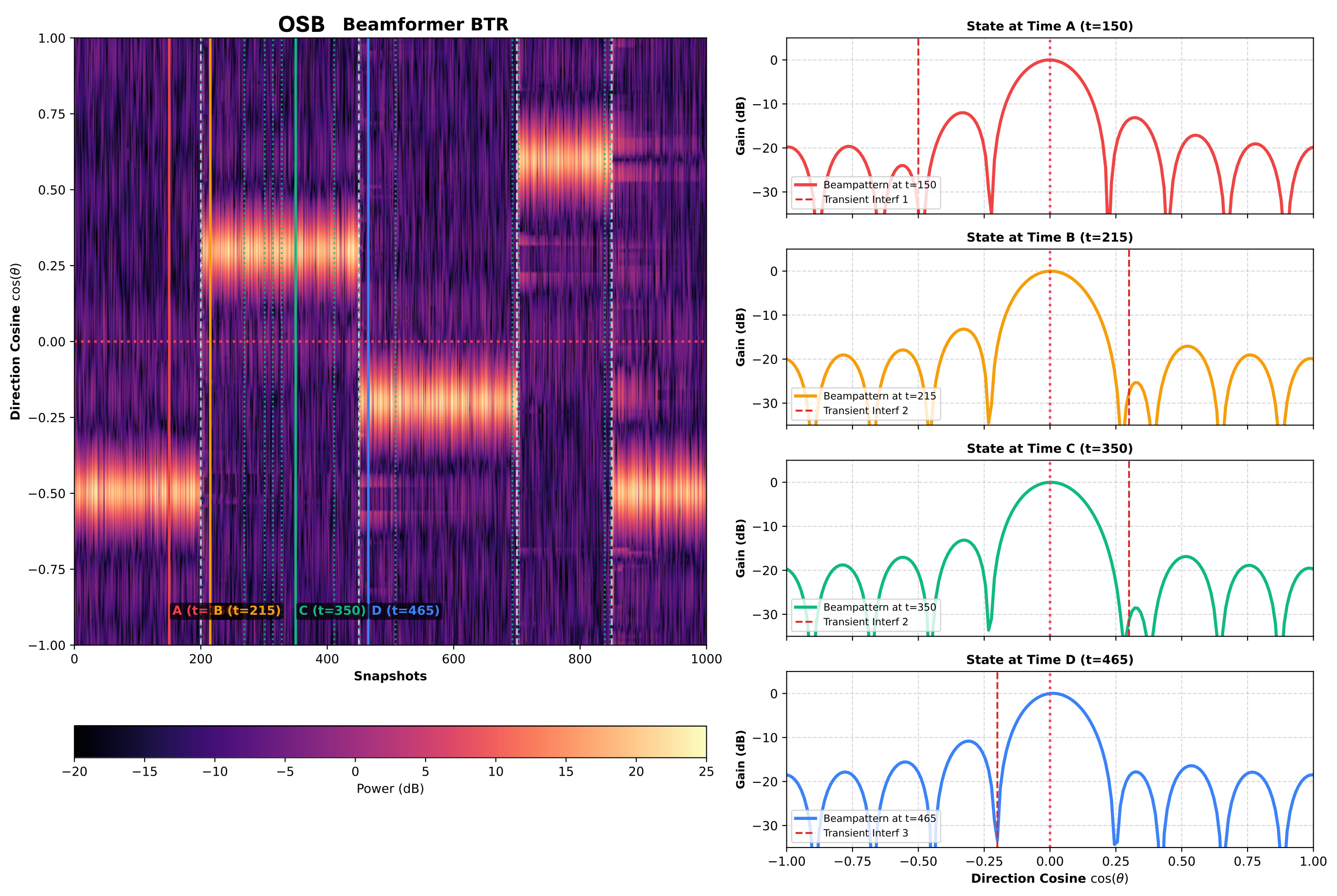}
    \caption{Demonstrative evaluation of the Online Segmented Beamformer (OSB). The visualization depicts the spatial-temporal BTR and highlights the OSB beampattern at marked analysis times. Notably, the OSB rapidly adapts to the changing environment and places nulls in the correct location.}
    \label{fig:demonstrative_example_beampatterns}
\end{figure}

To further inspect the transient response, Figure \ref{fig:demonstrative_example_beampatterns} extracts instantaneous beampatterns at four analysis times: $t \in \{150, 215, 350, 465\}$. The snapshots at $t=150$ and $t=350$ represent stable periods where the anchor window has integrated sufficient samples to form deep nulls in the direction of the active interferers. Conversely, the snapshots at $t=215$ and $t=465$ capture the algorithm immediately following the abrupt spatial shifts at $t=200$ and $t=450$. During these transient phases, the algorithm drops the previous spatial nulls and initiates new nulls corresponding to the updated interferer locations, preventing the suppression of inactive spatial sectors.

\subsection{Online Segmented Beamforming in Piecewise Constant Bearing Environments}

To isolate spatial transience, this simulation evaluates the algorithms in a piecewise constant bearing environment, testing the capacity to recover from sudden spatial shifts. The test involves a fixed pool of potential interferer locations, with the active spatial bearing changing abruptly. A ULA comprising $M = 15$ elements utilizes a standard half-wavelength spacing designed for a target signal frequency of $1000$ Hz with a speed of sound $c = 343$ m/s. The simulation length spans $T = 20,000$ snapshots. A target signal arrives continuously from broadside ($90^\circ$).

\begin{figure}[htbp]
\centering\includegraphics[width=\linewidth]{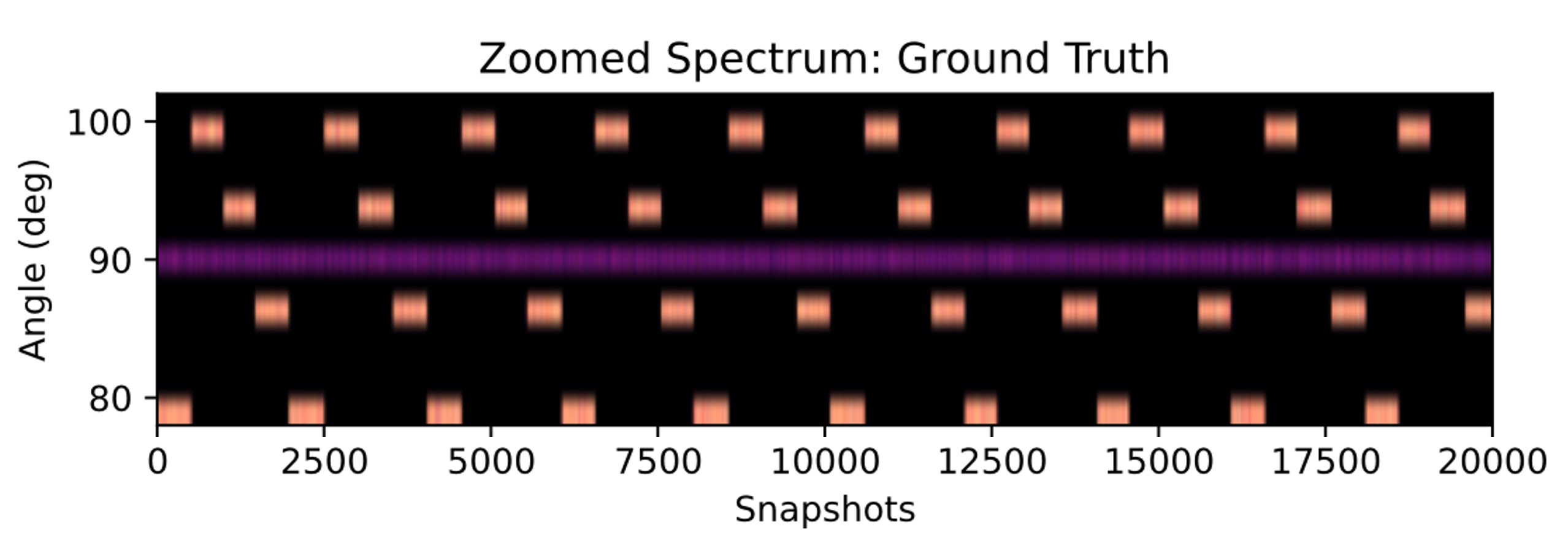}
\caption{A depiction of the bearing time record for the piecewise stationary in bearing simulation.}
\label{fig:pws_bearing_btr}
\end{figure}

The spatial environment contains a pool of $4$ potential interferer directions. To ensure challenging interference conditions without main-lobe contamination, these directions originate randomly from angular regions where the conventional beamformer's quiescent response provides between $4$ dB and $15$ dB of suppression. Exactly one of these interferers operates actively at any given time. The duration of each active block randomizes, consisting of a base length of $500$ snapshots perturbed by a uniform jitter of $\pm 30$ snapshots, resulting in true stationary segment lengths ranging from $470$ to $530$ snapshots. At the end of each block, the active interferer shifts to another location within the candidate pool. The target signal and interferers generate via complex Gaussian noise scaled such that the target SNR reaches $-5$ dB, while the active interferer arrives with an INR of $11$ dB. Additive white sensor noise operates with unit variance. Figure \ref{fig:pws_bearing_btr} depicts the bearing time record of the simulated environment.

\begin{figure}[htbp]
\centering\includegraphics[width=\linewidth]{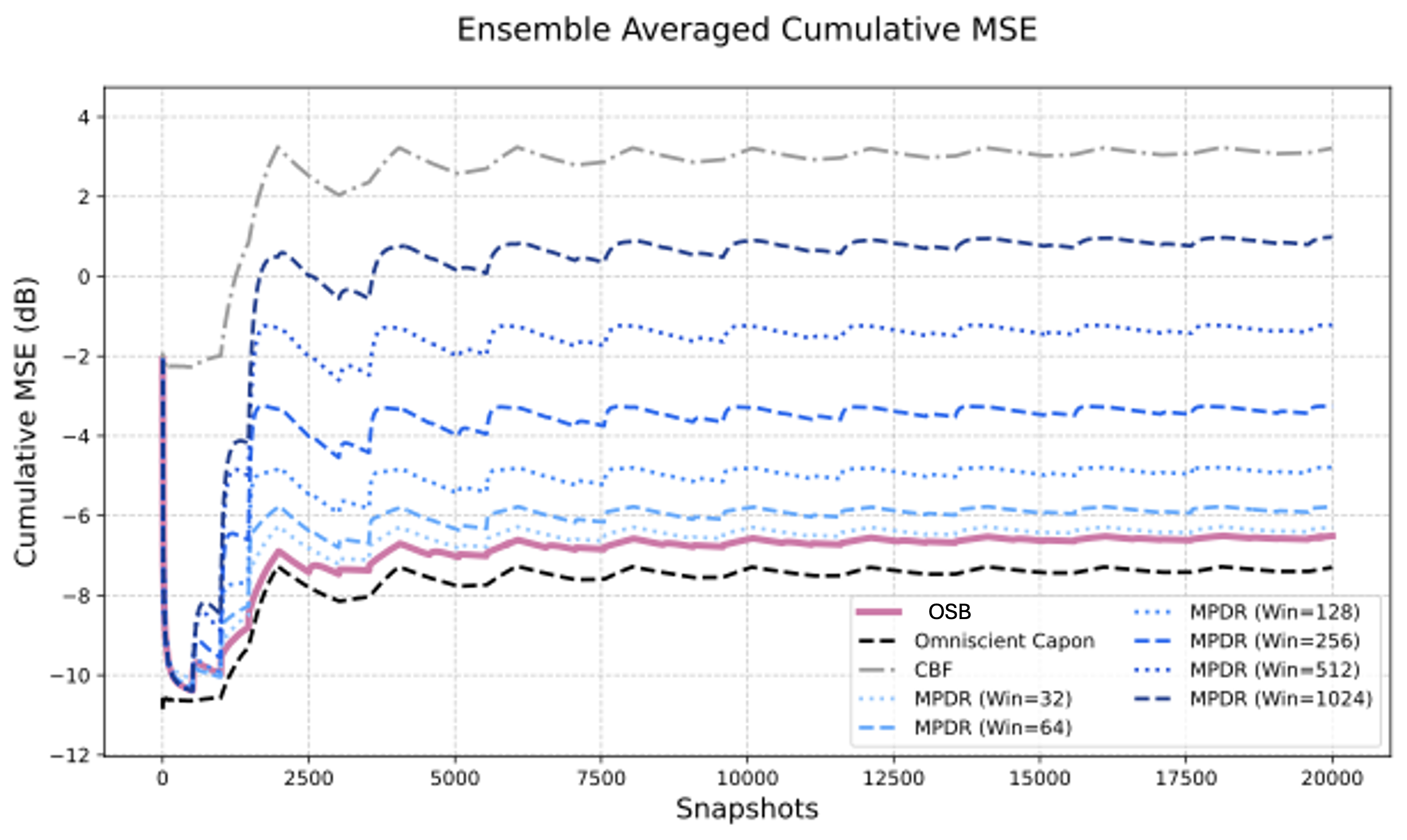}
\caption{The ensemble average mean-squared error of the algorithms in estimating the desired signal in a piecewise constant in bearing scenario. Lower values indicate better performance. The OSB matches the performance of the best sliding window.}
\label{fig:pws_bearing_mse}
\end{figure}

The OSB benchmarks against standard sliding-window MPDR beamformers, evaluated at exponentially spaced window lengths of $32$, $64$, $128$, $256$, $512$, and $1024$ snapshots. Additional baselines include an Omniscient Capon beamformer (utilizing the exact instantaneous covariance), and the Conventional Beamformer (CBF). All adaptive algorithms employ adaptive diagonal loading \cite{mittal2026adaptive}. Performance quantification relies on the cumulative Mean Squared Error (MSE) computed over $200$ Monte Carlo trials. Figure \ref{fig:pws_bearing_mse} demonstrates that the OSB tracks the optimal integration length. While short static windows exhibit rapid adaptation during the spatial jumps, they suffer from steady-state estimation variance. Conversely, long static windows provide steady-state suppression but fail completely during the abrupt bearing shifts. The segmented approach dynamically detects these boundaries, matching the transition performance of short windows while recovering the stability of long windows during the stationary blocks.

\subsection{Online Segmented Beamforming in Piecewise Constant Time Environments}

To challenge the assumption of constant stationarity timescales, this scenario introduces interference blocks of randomized durations. A ULA comprising $M = 15$ elements utilizes standard half-wavelength spacing designed for a target frequency of $1000$ Hz and a speed of sound $c = 343$ m/s. The timeline spans $T = 20,000$ snapshots, with a continuous target signal arriving from broadside ($90^\circ$) at an SNR of $-5$ dB.

\begin{figure}[htbp]
\centering\includegraphics[width=\linewidth]{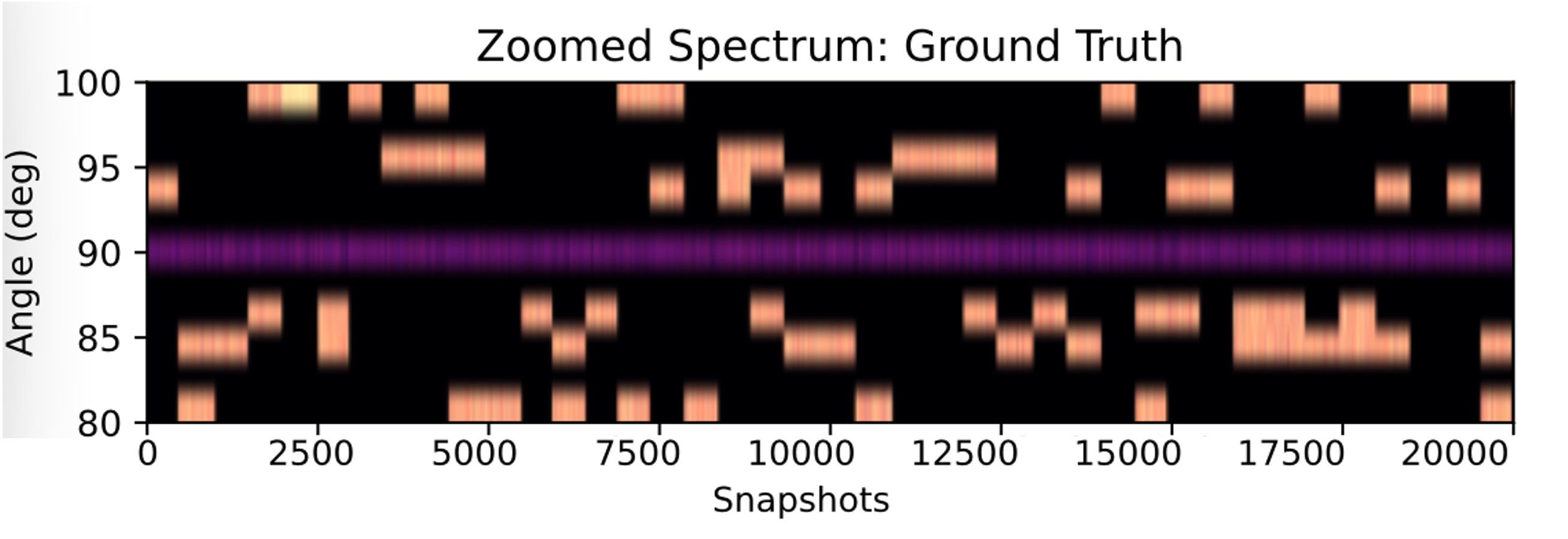}
\caption{A depiction of the bearing-time record for the piecewise constant time simulation for one trial.}
\label{fig:pws_time_btr}
\end{figure}

In this scenario, a pool of candidate interferer directions excludes the main lobe, selecting angles where the conventional beamformer provides between $3$ dB and $15$ dB of suppression. Exactly two independent interferers act simultaneously at any given moment. The active interferers draw randomly from the candidate pool, and each arrives with an Interference-to-Noise Ratio (INR) of $12$ dB. The duration of each active block randomizes. The base duration of $500$ snapshots experiences perturbation by a uniform jitter of $\pm 50$ snapshots, resulting in true stationary segment lengths that vary unpredictably between $450$ and $550$ snapshots. This variability explicitly breaks the assumption of a constant stationarity timescale. Additive white sensor noise operates with unit variance, and adaptive diagonal loading applies to all beamformers. Figure \ref{fig:pws_time_btr} depicts the bearing-time record.

\begin{figure}[htbp]
\centering\includegraphics[width=\linewidth]{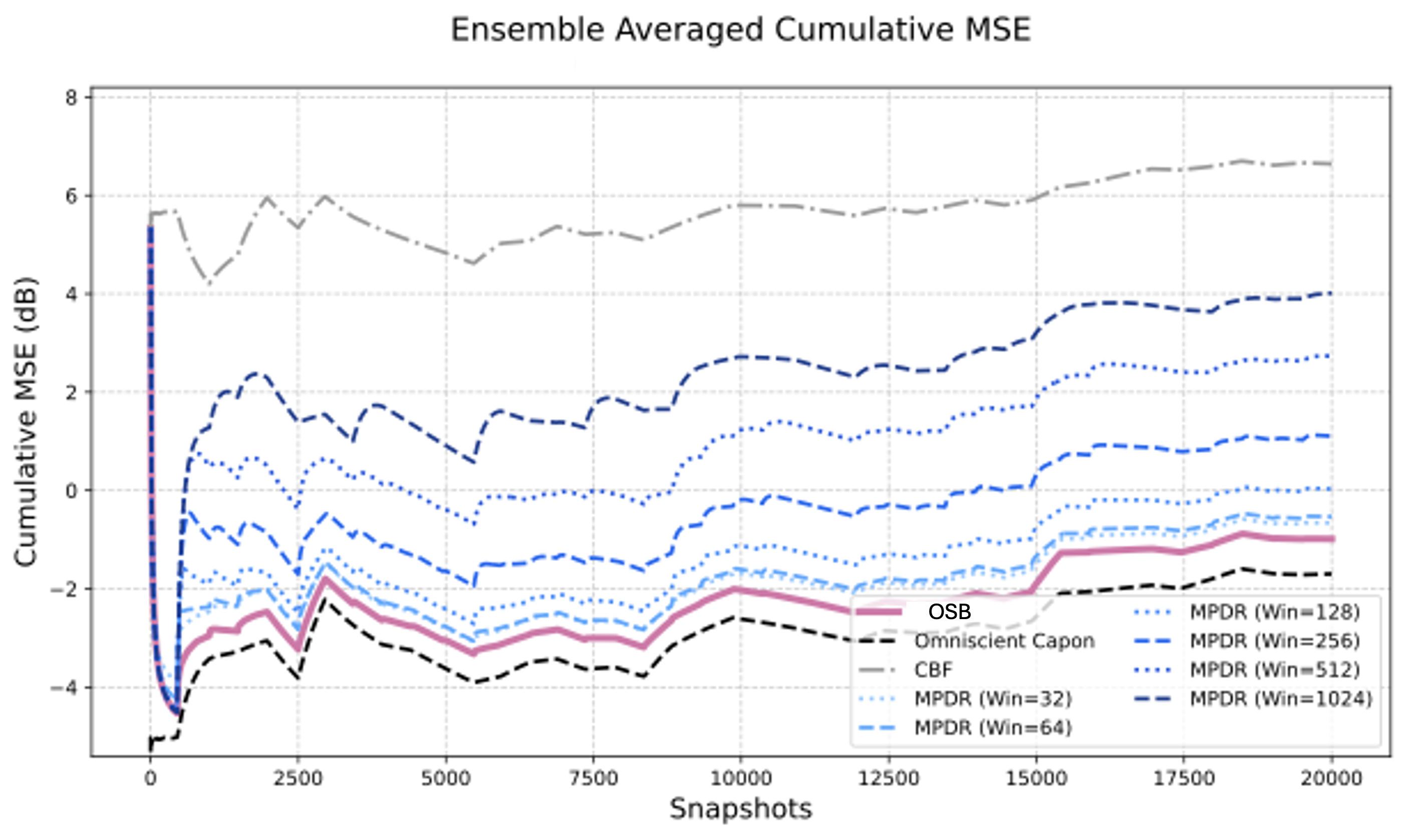}
\caption{The ensembled averaged mean-squared error of each of the algorithms in estimating the desired signal in the piecewise constant time scenario. The OSB outperforms the best sliding window chosen in hindsight.}
\label{fig:pws_time_mse}
\end{figure}

The OSB is benchmarked against fixed sliding-window MPDR beamformers with exponentially spaced memory lengths of $32$, $64$, $128$, $256$, $512$, and $1024$ snapshots. Additional baselines include the Omniscient Capon beamformer and the CBF. Figure \ref{fig:pws_time_mse} compares the cumulative mean-squared error (MSE) across $200$ Monte Carlo trials. The results demonstrate that fixed-memory sliding windows fail to achieve optimal performance across the fluctuating temporal blocks. A short window (e.g., $128$ snapshots) becomes overly responsive during the stable portions of the $550$-snapshot segments, whereas a long window (e.g., $1024$ snapshots) incurs penalties by aggressively averaging across multiple distinct interference blocks. The OSB consistently achieves a lower overall cumulative error. By continuously evaluating the local stationarity and greedily resetting the spatial covariance matrix at change points, the segmented framework dynamically adapts its effective memory length to the fluctuating duration of each interference state.

\subsection{Online Segmented Beamforming: Birth-Death Process}

To model continuous, stochastic environmental transitions, a Markov-driven birth-death interference process tests real-time regret-minimizing behavior. A ULA comprising $M = 15$ elements simulates with a standard half-wavelength inter-element spacing of $0.1715$ m, assuming a target signal frequency of $1000$ Hz and a speed of sound of $343$ m/s. The observation window spans $T = 20,000$ snapshots. A continuous target signal arrives from broadside ($90^\circ$) with a fixed Signal-to-Noise Ratio (SNR) of $-5$ dB. The interference environment evolves across a candidate pool of spatial angles, explicitly excluding the main lobe of the target signal. At any given time, a maximum of two independent interferers operate simultaneously, each arriving with an Interference-to-Noise Ratio (INR) of $12$ dB. The temporal evolution follows a Markov birth-death process: an inactive state transitions to generate a new active interferer with a probability of $p_{\text{birth}} = 0.02$, yielding an expected idle ("off") time of $50$ snapshots before a new interference onset. Conversely, any currently active interferer disappears with a probability of $p_{\text{death}} = 0.001$, yielding an expected active duration ("on" time) of $1000$ snapshots per interferer.

\begin{figure}[htbp]
\centering\includegraphics[width=\linewidth]{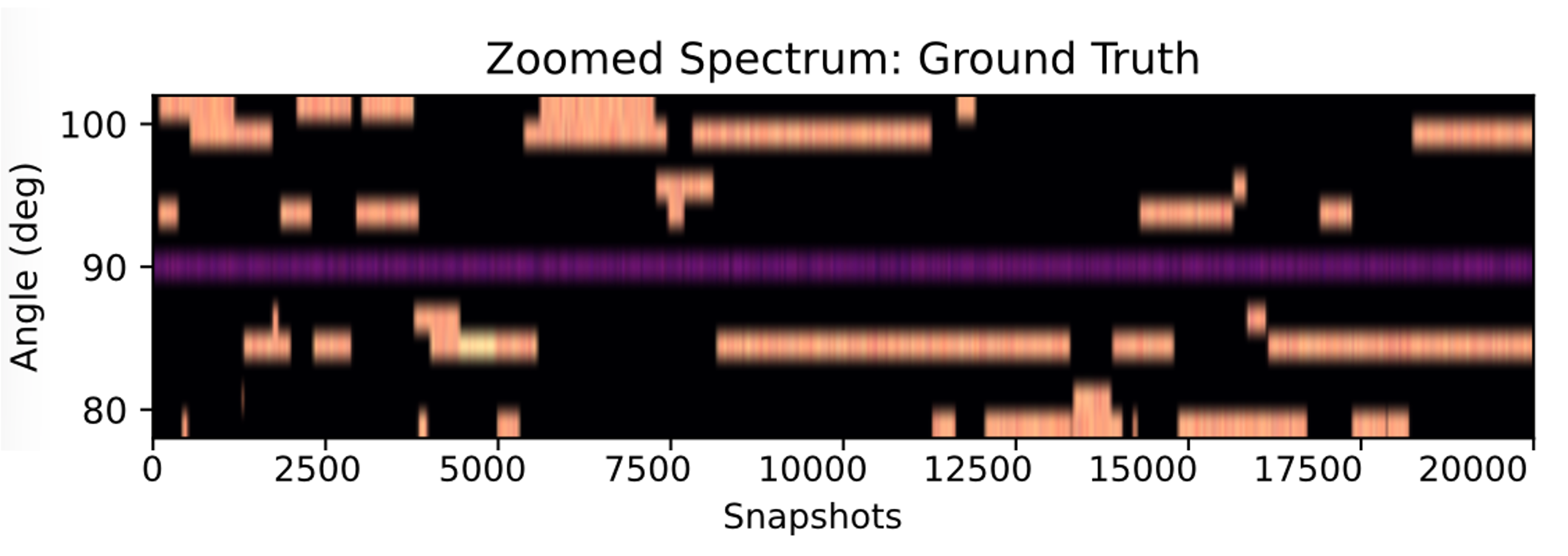}
\caption{The ground truth spatial spectrum output across the angular manifold over time for the birth-death simulation.}
\label{fig:birth_death_simulation_plots.png}
\end{figure}

The OSB benchmarks against a suite of adaptive spatial filters over $200$ Monte Carlo trials. Standard sliding-window Minimum Power Distortionless Response (MPDR) beamformers evaluate across fixed memory lengths: $32$, $64$, $128$, $256$, $512$, and $1024$ snapshots. An Omniscient Capon bound serves as the baseline. All adaptive algorithms utilize a adaptive diagonal loading to ensure matrix invertibility \cite{mittal2026adaptive}. For the OSB, the complexity penalty defined as $C = 4.8$, and the minimum segment length bounds at $\tau = 5$ snapshots. Figure \ref{fig:birth_death_simulation_plots.png} shows the scanned response for the methods.

\begin{figure}[htbp]
\centering\includegraphics[width=\linewidth]{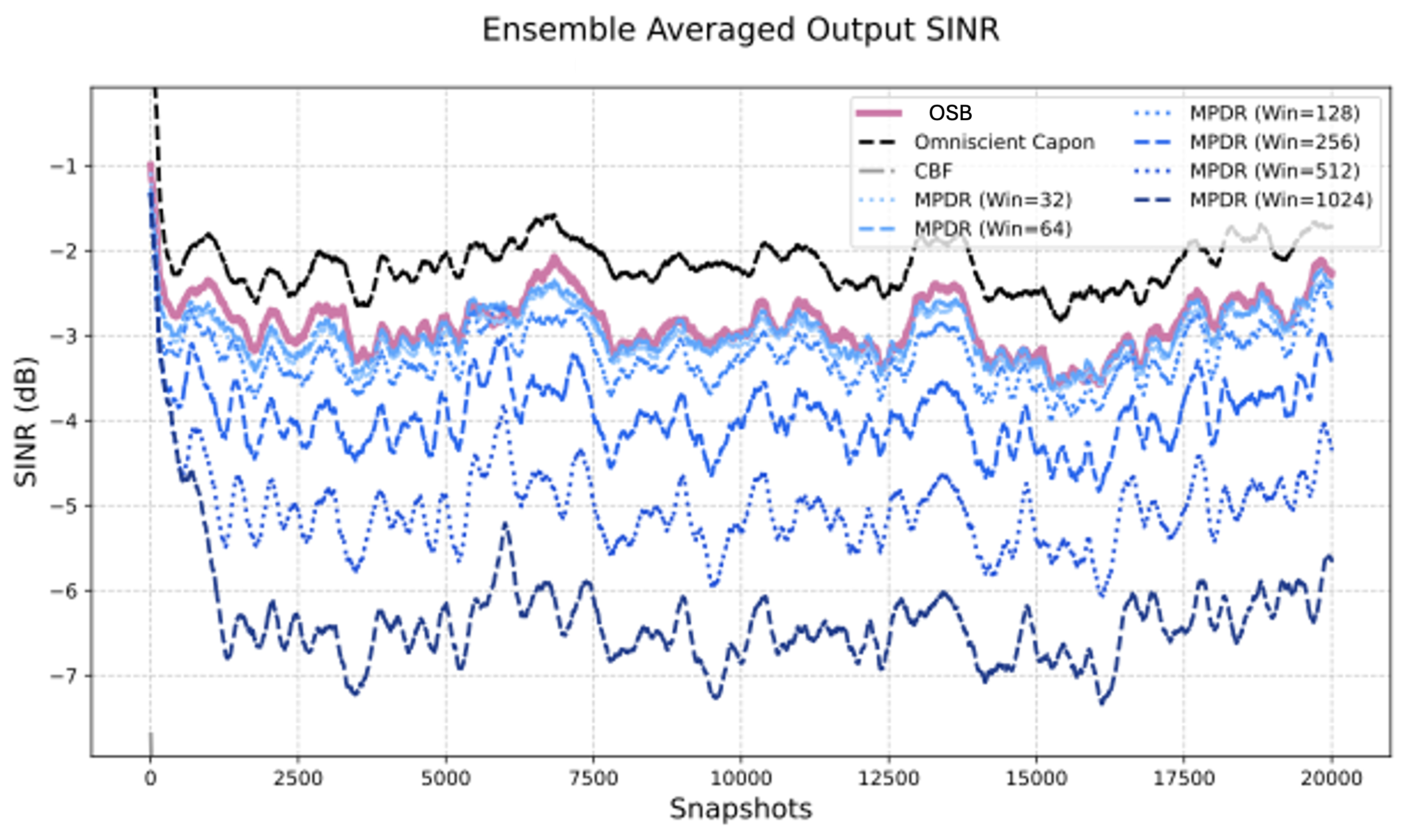}
\caption{Reconstructed Target SINR over time. The OSB method rapidly converges during interference state changes.}
\label{fig:birth_death_sinr_sim.png}
\end{figure}

\begin{figure}[htbp]
\centering\includegraphics[width=\linewidth]{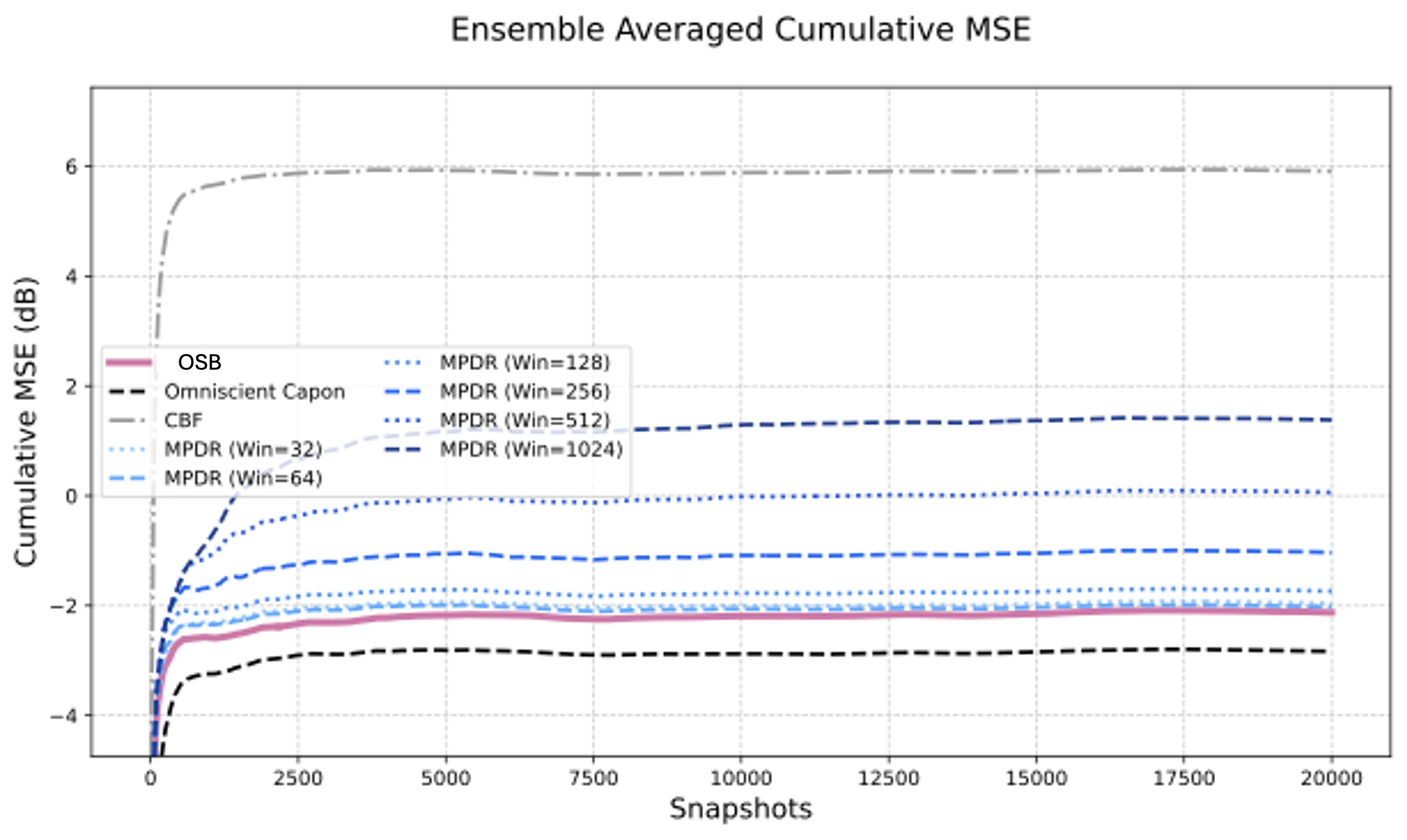}
\caption{Ensemble averaged cumulative Mean Squared Error (MSE) of the target signal estimate for the birth-death scenario. The OSB matches or exceeds the performance of the best fixed-window baseline.}
\label{fig:birth_death_mse_sim.png}
\end{figure}

The evaluation quantifies performance using the output Signal-to-Interference-plus-Noise Ratio (SINR), shown in Figure \ref{fig:birth_death_sinr_sim.png}, and the cumulative Mean Squared Error (MSE), shown in Figure \ref{fig:birth_death_mse_sim.png}. The results indicate that fixed-memory sliding windows suffer from an inherent bias-variance tradeoff. In contrast, the OSB detects abrupt changes in the acoustic scene and instantaneously resets the covariance estimate. This hard switching behavior minimizes the transient error following an interference jump, allowing the algorithm to rapidly converge to the new optimal weight vector and yield a lower steady-state error. The regret-minimizing behavior demonstrates that as more samples arrive, the difference in cumulative error between the best possible retrospective beamformer and the online method converges toward zero.

\section{Experiments}

\subsection{SwellEx-96 HLA}

To evaluate the operational capacity of the OSB in physical, unmodeled oceanic conditions, this experiment processes real-world hydrophone data from the SwellEx-96 Shallow Water Acoustic Experiment. Specifically, it utilizes the South Horizontal Line Array (HLA) recording. The array consists of 28 elements distributed in a planar geometry at a constant depth of 198 meters.

The acoustic data, sampled at $3276.8$ Hz, processes over a 10-minute segment beginning at minute 34 of the recording. The time-domain signals segment into snapshots of 512 samples with a 50 percent overlap. The analysis focuses on a single narrowband frequency component at $49$ Hz. Two-dimensional steering vectors, denoted by $\boldsymbol{\nu}$, compute across an azimuthal grid spanning $-90^\circ$ to $90^\circ$ in $1^\circ$ increments, assuming a nominal sound speed of $1500$ m/s. To ensure numerical stability within the recursive covariance updates, the frequency-domain snapshots receive global normalization by their standard deviation prior to spatial filtering. 

\begin{figure}[htbp]
    \centering
    \includegraphics[width=\linewidth]{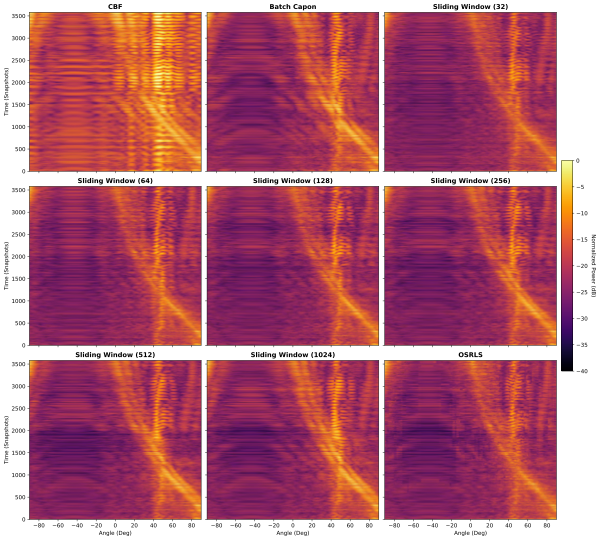}
    \caption{Scanned Response of the various algorithms at 49Hz in the SwellEx-96 dataset. The OSB matches the performance of the best sliding window without relying on fixed a priori temporal assumptions.}
    \label{fig:SwellEx Scanned Response}
\end{figure}

The OSB, with a maximum search limit $K=60$, evaluates against several baseline spatial filters: the Conventional Beamformer (CBF), the static batch Capon (MVDR) beamformer, and standard sliding-window MPDR beamformers configured with memory lengths of 32, 64, 128, 256, 512, and 1024 snapshots. All methods use adaptive diagonal loading \cite{mittal2026adaptive}. For the OSB, the complexity penalty defines as $C = 0.1$, the minimum segment length stands at $\tau = 1$ snapshot. The bearing time record for each of the considered methods appears in Figure \ref{fig:SwellEx Scanned Response}.

\begin{figure}[htbp]
    \centering
    \includegraphics[width=\linewidth]{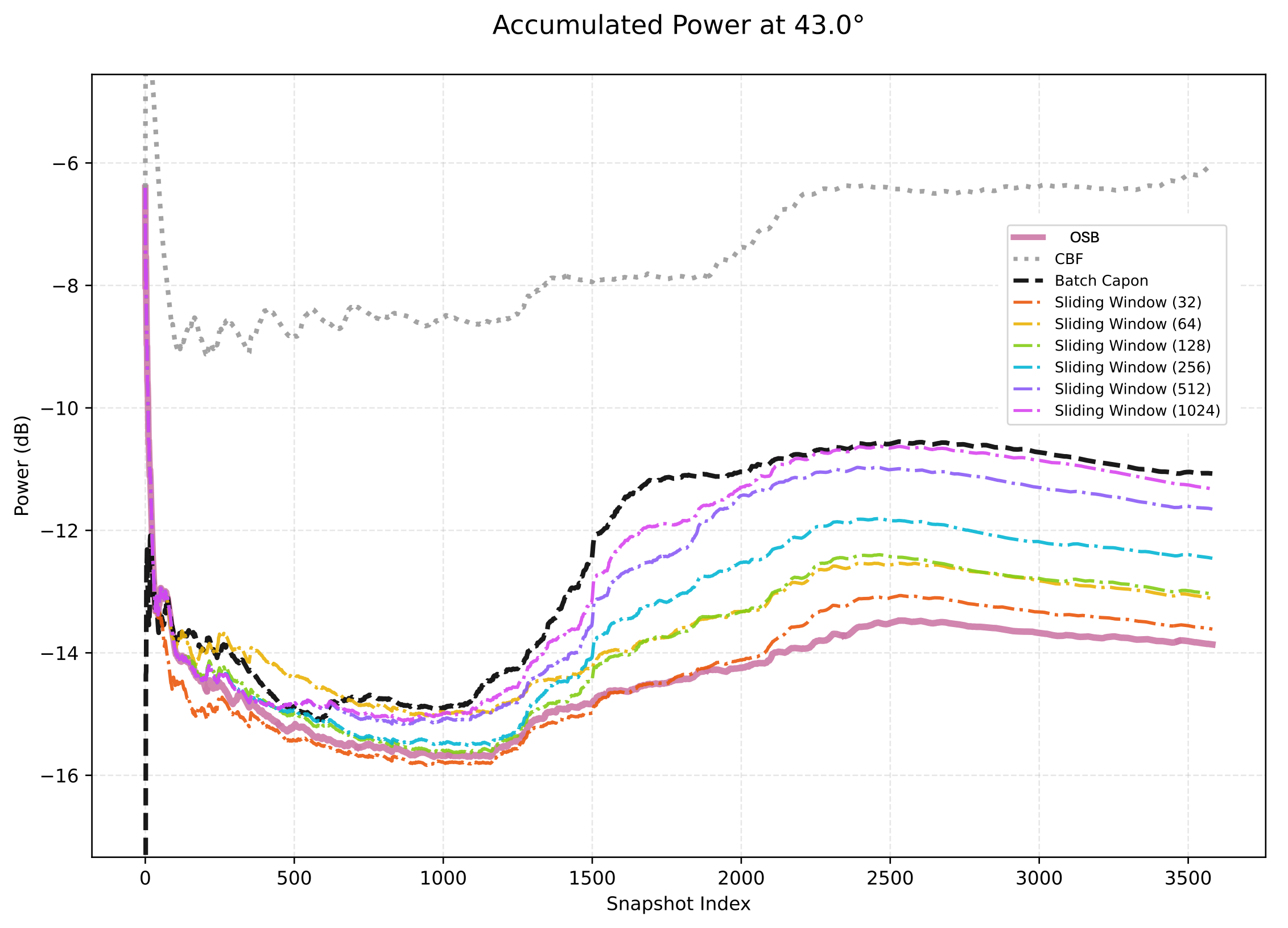}
    \caption{Accumulated output power at $43^\circ$. The OSB minimizes its regret over time, performing on par with the best sliding window MPDR.}
    \label{fig:SwellEx power}
\end{figure}

\begin{figure}[htbp]
    \centering
    \includegraphics[width=\linewidth]{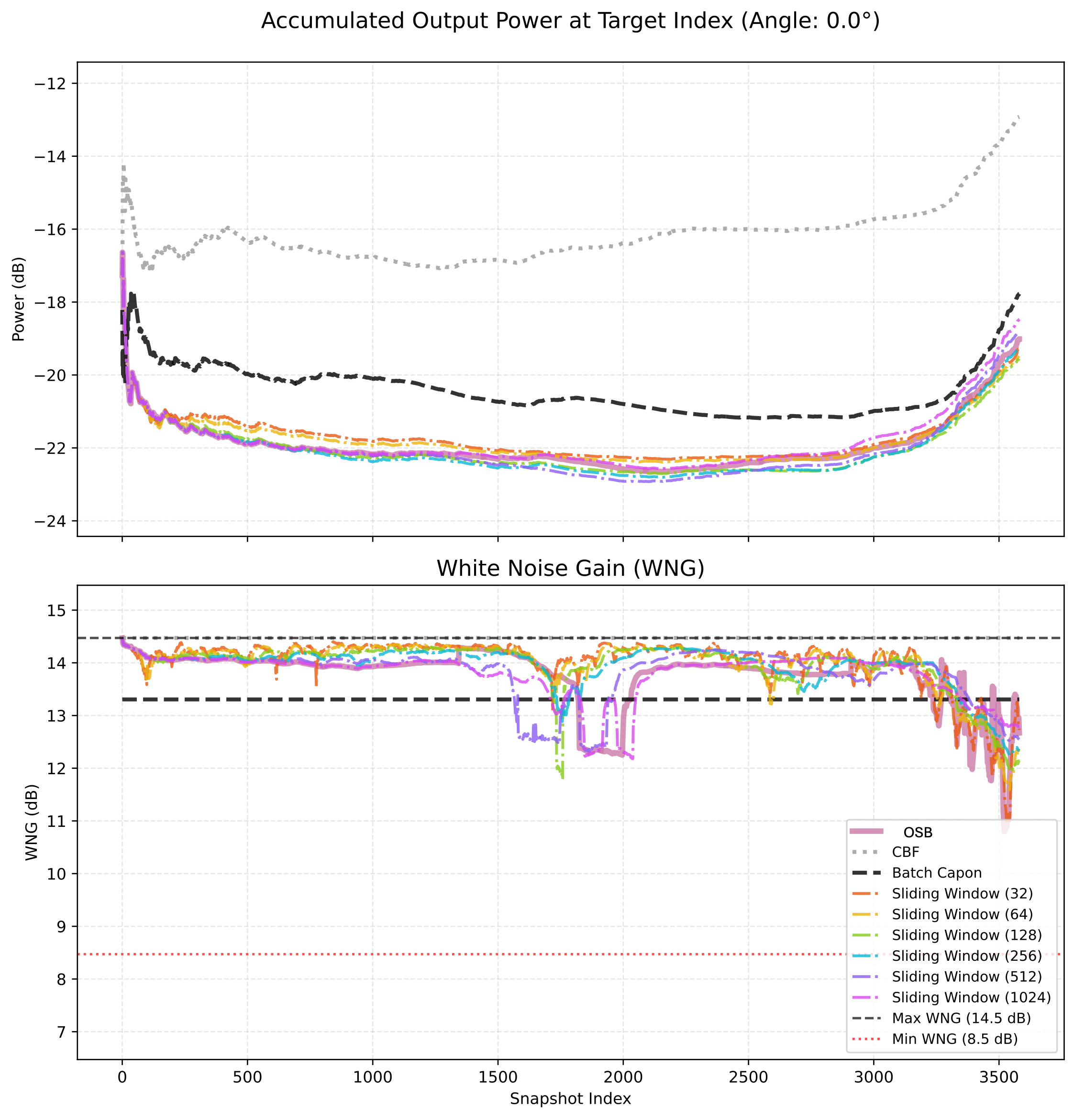}
    \caption{Accumulated output power and white noise gain at $0^\circ$. The OSB performs reliably across the target metric dimensions.}
    \label{fig:SwellEx WNG}
\end{figure}

Performance quantification relies on evaluating the cumulative output power of the beamformers over the observation period. The accumulated power extracts specifically at a fixed steering angle of $43^\circ$, which corresponds to the angle of the source with constant bearing, and at $0^\circ$. Figure \ref{fig:SwellEx power} shows the accumulated power at $43^\circ$. The segmented method eventually performs comparably with the best sliding window beamformer. Figure \ref{fig:SwellEx WNG} compares the white noise gain of the different beamformers and the accumulated output power at $0^\circ$. The analysis of the experimental data confirms that the segmented approach dynamically manages the bias-variance tradeoff inherent in the sample covariance matrix estimation without requiring a fixed prior on the environmental coherence time.

\subsection{Distributed Microphone Array Dataset}

To assess performance in a physical enclosed space with commercial microphone hardware, this evaluation applies the framework to multi-talker human speech recordings from the Massive Distributed Microphone Array Dataset \cite{corey_massive_2019}. This dataset contains acoustic measurements captured by 160 microphones distributed throughout a conference room measuring 13 m by 9 m, with a reverberation time ($T_{60}$) of approximately $800$ ms. 

The sensing infrastructure comprises two types of arrays: wearable microphone arrays containing 16 sensors each and tabletop arrays containing 8 sensors each. The evaluation utilizes a spatially distributed subarray created by selecting 40 channels uniformly from the available microphone positions. This subset mimics an ad-hoc distributed acoustic sensor network, lacking uniform geometry and presenting real-world multi-path effects.

The target and interference signals consist of 60-second continuous speech clips derived from the VCTK corpus, played simultaneously through multiple loudspeakers to simulate a dynamic cocktail party scenario. Ambient background noise recorded in the conference room supplements the mixtures to ensure realistic signal-to-noise ratios.

The experimental design restricts the acoustic scene to four active talkers distributed spatially across the room. The beamforming task requires extracting the desired talker while continuously adapting to suppress the remaining three interfering talkers and the ambient room noise. The Relative Transfer Functions (RTFs) derive from isolated exponential frequency sweep recordings provided in the dataset for each loudspeaker location.

The OSB algorithm processes the 40-channel mixture in the STFT domain using a frame size of 1024 samples and 50\% overlap. The framework benchmarks against the standard MPDR beamformer utilizing fixed sliding windows of varying lengths.

\begin{figure}[htbp]
\centering
\includegraphics[width=\linewidth, height = 15cm]{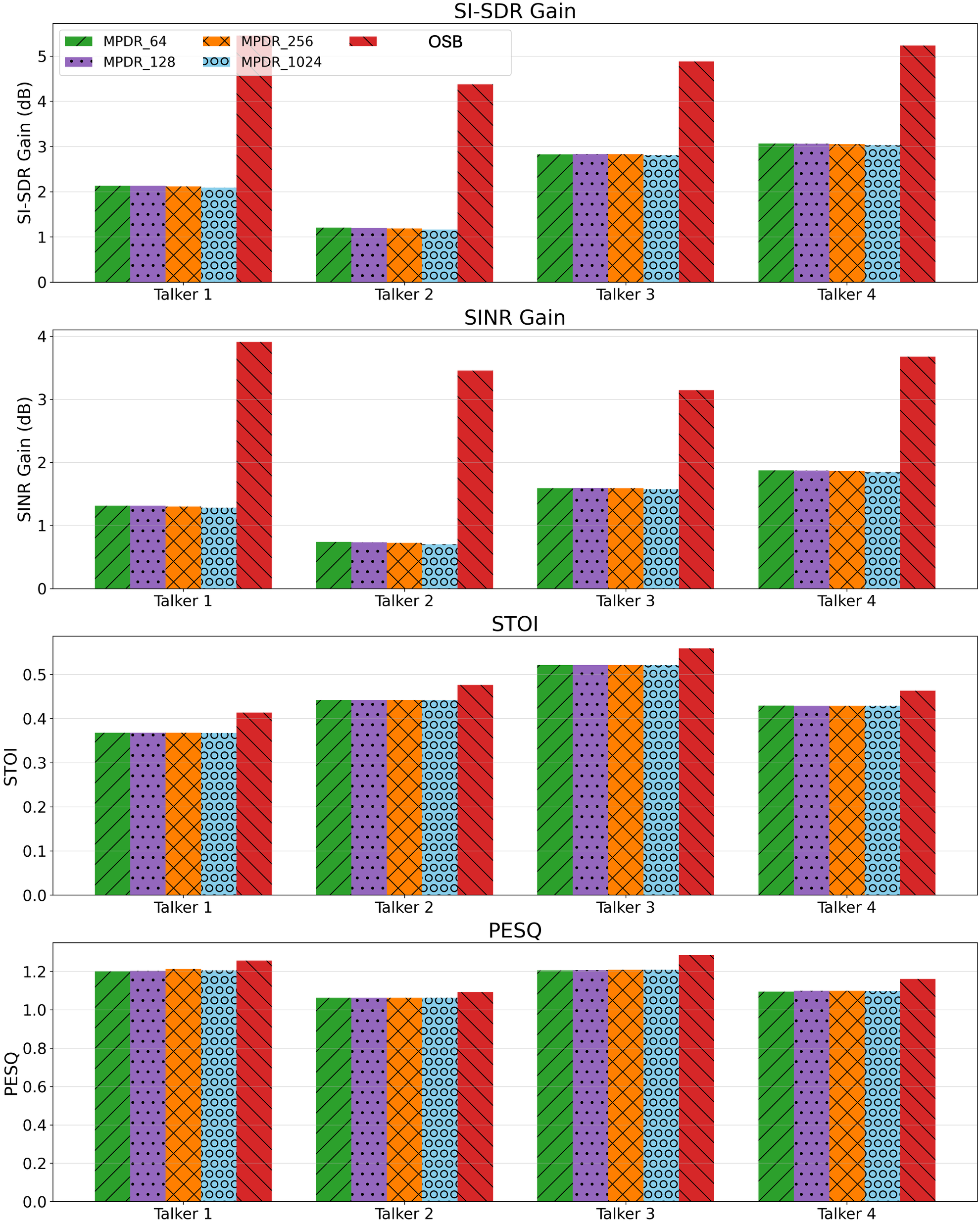} 
\caption{Performance comparison of the OSB versus standard MPDR across different target speakers in the Massive Microphone Array Dataset. OSB results in better performance across all tested metrics. }
\label{fig:performance_plot}
\end{figure}

As shown in Figure \ref{fig:performance_plot}, the real-world reverberant data corroborates the simulation findings. The fixed-window MPDR formulations struggle to balance the bias-variance tradeoff; short windows suffer from covariance estimation noise due to the lack of spatial averaging, while long windows fail to track phase shifts and multi-path reflections inherent in a physical room. The OSB autonomously partitions the observation timeline, resulting in consistent SI-SDR and PESQ metrics. By updating covariance boundaries dynamically, the method maintains spatial nulls against the competing talkers without distorting the target speech.

\section{Conclusion}

This work presents a principled framework for adaptive beamforming in non-stationary acoustic environments. By identifying the fundamental limitation of standard adaptive beamformers, namely the reliance on fixed integration windows that smear distinct acoustic states, the beamforming problem reformulates as a joint task of estimation and temporal segmentation.

Drawing on the theory of Segmented Least Squares, the analysis derives the Segmented Beamformer as a formal generalization of the classic Capon estimator. The formulation demonstrates that the standard MVDR solution represents a specific, single-segment instance of this broader class of estimators. By applying Bellman’s principle of optimality, the batch formulation guarantees a globally optimal partitioning of the observation record. This ensures that covariance matrices estimate strictly over intervals of true stationarity, thereby maximizing the available degrees of freedom for interference suppression.

To bridge the gap between theoretical optimality and practical real-time processing, the framework details the Online Segmented Beamformer. This algorithm functions as a universal estimator, dynamically adapting its effective memory length to the underlying rate of environmental change. A theoretical analysis proves that this greedy online strategy achieves a cumulative regret that grows only logarithmically relative to the optimal batch segmentation chosen in hindsight.

Simulations and experimental validations on the Microphone Array and SwellEx-96 datasets demonstrate the robustness of the framework. The Segmented Beamformer consistently matches or outperforms fixed-memory baselines, including RLS and sliding-window Capon filters, in dynamic scenarios. By allowing the statistical structure of the data to intrinsically dictate the adaptation rate, this framework provides a parameter-free alternative to traditional adaptive beamforming, suited for time-varying acoustic applications ranging from speech processing to underwater sonar arrays.


\bibliographystyle{IEEEtran}
\bibliography{ieee-bibliography}

\end{document}